\documentclass{emulateapj}
\usepackage{graphicx}
\usepackage{epstopdf}
\usepackage{amsmath}

\begin{document}
\title{The Ties that Bind?  Galactic Magnetic Fields and Ram Pressure Stripping}
\author{Stephanie Tonnesen$^1$ and James Stone$^2$}
\affil{Department of Astrophysics, Princeton University, Peyton Hall, Princeton, NJ, 08544}
\email{1 stonnes@astro.princeton.edu\\
2 jstone@astro.princeton.edu}




\begin{abstract}
One process affecting gas-rich cluster galaxies is ram pressure stripping, i.e. the removal of galactic gas through direct interaction with the intracluster medium.  Galactic magnetic fields may have an important impact on the stripping rate and tail structure.  We run the first magnetohydrodynamic simulations of ram pressure stripping that include a galactic magnetic field, using 159 pc resolution throughout our entire domain in order to resolve mixing throughout the tail.  We find very little difference in the total amount of gas removed from the unmagnetized and magnetized galaxies, although a magnetic field with a radial component will initially accelerate stripped gas more quickly.  In general, we find that magnetic fields in the disk lead to slower velocities in the stripped gas near the disk and faster velocities farther from the disk.  We also find that magnetic fields in the galactic gas lead to larger unmixed structures in the tail.  Finally, we discuss whether ram pressure stripped tails can magnetize the ICM.  We find that the total magnetic energy density grows as the tail lengthens, likely through turbulence.  There are $\mu$G-strength fields in the tail in all of our MHD runs, which survive to at least 100 kpc from the disk (the edge of our simulated region), indicating that the area-filling factor of magnetized tails in a cluster could be large.  \end{abstract}


\section{Introduction}

As galaxies orbit within a cluster, their interstellar medium (ISM) may interact directly with the intracluster medium (ICM), the hot halo of gas bound by the cluster gravitational potential.  This type of interaction may take the form of ram pressure stripping (RPS), in which the ISM is removed through momentum transfer by the ICM (Gunn \& Gott 1972).  ISM-ICM interactions also include continuous stripping by thermal evaporation, Kelvin-Helmholtz instabilties, or turbulent viscous stripping (Cowie \& Songaila 1977; Chandrasekhar 1961; Nulsen 1982). 

ISM-ICM interactions have been well-studied using hydrodynamic simulations (e.g. Abadi et al. 1999; Schulz \& Struck 2001; Roediger \& Hensler 2005; Roediger \& Br{\"u}ggen 2006; Roediger \& Br{\"u}ggen 2007;  Quilis, Moore \& Bower 2000; Kronberger et al. 2008; Kapferer et al. 2009; J{\'a}chym et al. 2009; Tonnesen \& Bryan 2009, 2010, 2012; Tonnesen et al. 2011; sticky particle simulations in e.g. Vollmer et al. 2001; Vollmer et al. 2002).  Work including radiative cooling has found that stripped tails can extend several hundred kpc (Kronberger et al. 2008; Kapferer et al. 2009; Tonnesen \& Bryan 2010; Tonnesen et al. 2011), in agreement with HI observations of some cluster galaxies (e.g. Oosterloo \& van Gorkom 2005). 

However, very little work has been done using magnetohydrodynamic (MHD) simulations.  Ruszkowski et al. (2014) include ICM magnetic fields and radiative cooling, and find that intracluster magnetic fields affect the tail morphology.  The clumpy gas in a hydrodynamic radiatively cooling tail is smoothed to a more filamentary structure when intracluster magnetic fields are included.  Pfrommer \& Dursi (2010) argue that their MHD simulations show that observations of polarized intensity contours on the face of ram pressure stripped galaxies can be used to determine the orientation of magnetic fields in clusters.  These simulations of RPS only include magnetic fields in the ICM and neglect galactic magnetic fields. 

Including magnetic fields in the galaxy and the ICM is important to understanding the physics behind observations of ram pressure stripped galaxies.  For example, enhancement of radio continuum emission has been observed in cluster galaxies, as well as an enhanced, though still tight, radio-to-far infrared (FIR) correlation (Gavazzi 1991; Niklas et al. 1995; Andersen \& Owen 1995).  Because the radio is enhanced relative to the FIR emission, Scodeggio \& Gavazzi (1993) and Rengarajan et al. (1997) argue that the increase in radio continuum emission cannot be entirely explained by enhanced star formation in cluster galaxies, and claim that therefore magnetic field compression by the ICM is also likely to be at work.

Using spatial information, Murphy et al. (2009) compare maps of the FIR-radio correlation between Virgo and normal galaxies.  Based on the FIR emission and expectations from field galaxies, they find radio deficits in galaxies undergoing ram pressure stripping along the face of the interaction between the galaxy and the ICM.  In agreement with earlier works, they find that cluster galaxies have enhanced global radio emission, but they are the first to connect this enhanced global emission with local radio deficits in ram pressure stripped galaxies.  This was then seen and discussed in the context of detailed multi-wavelength observations of several Virgo galaxies (Vollmer et al. 2009; 2010; 2013).  The scenario proposed by these authors is that low density gas and its associated magnetic fields and relativistic electrons are more easily stripped.  The observed enhanced radio emission is produced when mini-shocks accelerate cosmic rays in the ISM.  On the other hand, Pfrommer \& Dursi (2010) argue that magnetic draping of the intracluster magnetic field can explain the radio deficit observations. 

Including galactic magnetic fields may also be important in the study of stripped tails.  In fact the first stripped tails were observed in radio continuum emission (Gavazzi \& Jaffe 1987), indicating magnetic fields in the stripped gas.  Recently Sun et al. (2006; 2007) found a ram pressure stripped ``double-tail" in X-ray emission, which they argue may be due to confinement from magnetic fields.  Zhang et al. (2013) report on another double-tailed stripped galaxy in the same cluster (Ruszkowski et al. 2014 report that they reproduce a double tail with an intracluster magnetic field, and intend to determine the brightness of this feature in future work).  

In addition, stripped tails have the same metallicity as their parent galaxy, and thus may be important to the enrichment of the ICM (this has been observed in, e.g. Kenney et al. 2014).  It is still under debate whether, in general, the ICM has a smooth, flat distribution of metals beyond a central cD-dominated peak (e.g. Werner et al. 2013; Leccardi \& Molendi 2008; Fujita et al. 2008; Ezawa et al. 1997); metallicities that, beyond the cD-dominated central region, fall slowly with increasing radius (e.g. Finoguenov et al. 2000; Baldi et al. 2007; Matsushita 2011); or if the ICM metallicity fluctuates on both small and large scales, as observed in A3667 (Lovisari et al. 2009).  Fluctuations in the ICM metallicity indicate recent enrichment, and RPS could then be an important enrichment mechanism (Domainko et al. 2006).  Including magnetic fields is important to understanding the rate at which stripped material mixes with the ICM.  

The magnetic field in the tail may also be important, as it may help to magnetize, or maintain magnetic fields in, the ICM (for a discussion of possible sources of the intracluster magnetic field see Brandenburg \& Subramanian 2005).  Several observations indicate that intracluster magnetic fields are $\sim$$\mu$G, possibly rising to nearly 10 $\mu$G in cluster centers (e.g. Pratley et al. 2013; Feretti et al. 2012; Bonafede et al. 2009; Murgia et al. 2009; Guidetti et al. 2008; Govoni et al. 2006; Govoni \& Feretti 2004; Clarke, Kronberg \& B{\"o}hringer 2001; Eilek \& Owen 2002; Clarke 2004; Vogt \& En{\ss}lin 2003, 2005).  Clarke, Kronberg \& B{\"o}hringer (2001) find that within about 500 kpc from the cluster center the area-covering factor of magnetic fields is about unity.  Subramanian et al. (2006) calculate using a simple analytical model that while galaxy wakes fill a small fraction of the volume of a cluster they could have an area covering factor near unity.  Arieli et al. (2011) use a cosmological simulation in which they assume a constant 3 $\mu$G galactic magnetic field and analytically model stellar outflows and ram pressure stripping from galaxies to study whether galactic gas can magnetize the ICM.  The authors find that without allowing for field amplification or dissipation, winds and stripping can magnetize the ICM to an average field strength of 0.9 $\mu$G in the central 100 kpc.  Studying the magnetic field strength and dissipation rate in tails in detail is necessary to determine the level to which stripped tails can magnetize the ICM.

In this paper, we run a set of high resolution simulations to understand the effect of galactic magnetic fields on ram pressure stripped disks and tails.  Here we focus on the effects of galactic magnetic fields, and therefore we do not include added physics such as radiative cooling or conduction.  Running an ideal MHD simulation is the first step in constraining the physical processes at work in the ICM that can affect stripped tails.  These constraints will come through comparisons of the field strength and structure in simulated and observed stripped tails, and in the mixing rate of stripped gas with the ICM.  This is a first step because non-ideal plasma transport processes such as anisotropic viscosity and conduction could have important effects on the mixing rate of the stripped tail of gas, the energy dissipation rate, and the propagation and decay of turbulence in the tail (Braginskii 1965; Lyutikov 2007, 2008; Schekochihin et al. 2009). 

The paper is structured as follows.  After a brief introduction to our methodology, we provide the details of our galaxy model in Section \ref{sec:galaxy}, and of the ICM in Section \ref{sec:ICM}.  We then discuss whether and how including a galactic magnetic field affects the disk and tail (\S \ref{sec:results}), first focusing on how magnetic fields affect the stripping rate in the disk (\S \ref{sec:gasdisk}) and then how they affect the stripped tail (\S \ref{sec:gastail}).  In Section \ref{sec:BICM} we focus directly on the magnetic field in the stripped tail, and discuss to what extent ram pressure stripped tails can add to the magnetization of the ICM.  In Section \ref{sec:discussion} we compare our results to previous simulations and observations.  Finally, we conclude in \S \ref{sec:conclusion} with a summary of our results and discuss future work.

\section{Methodology}

We use the MHD grid code Athena (Stone et al. 2008).  To follow the gas, we employ a mesh for solving the fluid equations including gravity.  In our implementation we use a single grid throughout the box with a cell size of 159 pc.  Much of the post-processing analysis of these simulations was performed using yt, an open-source analysis toolkit (Turk et al. 2011). 

\subsection{The Galaxy}\label{sec:galaxy}

Our galaxy is placed at a position corresponding to (83,83,52) kpc from the corner of our (166,166,151) kpc computational box (in one of our runs, described below and labeled DIP, the box is slightly narrower, with the disk (68,68,62) kpc from the corner of our (136,136,160) kpc domain), so that we can follow the stripped gas for about 100 kpc.  The galaxy remains stationary throughout the runs.  The ICM wind flows along the z-axis in the positive direction, with the lower z boundary set for inflow and upper z boundary set as outflow.  The x and y boundaries are set to outflow in all four runs.

We describe our disk in detail in Tonnesen \& Bryan (2009; 2010), but repeat the salient points here. We choose to model a massive spiral galaxy with a flat rotation curve of 200 km s$^{-1}$.  It consists of a gas disk that is followed using the mesh algorithm (excluding self-gravity), as well as the static potentials of the (pre-existing) stellar disk, stellar bulge, and dark matter halo.  We directly follow Roediger \& Br\"uggen (2006) in our modeling of the stellar and dark matter potential and gas disk.  In particular, we model the stellar disk using a Plummer-Kuzmin disk (see Miyamoto \& Nagai 1975), the stellar bulge using a spherical Hernquist profile (Hernquist 1993), and the dark matter halo using the spherical model of Burkert (1995).  This dark matter halo model is compatible with observed rotation curves (Burkert 1995; Trachternach et al. 2008).  The equation for the analytic potential is equation (2) in Mori \& Burkert (2000).  Our stellar disk has a radial scale length of 4 kpc, a vertical scale length of 0.25 kpc and a total mass of 10$^{11}$ M$_{\odot}$; the stellar bulge has a scale length of 0.4 kpc and a total mass of 10$^{10}$ M$_{\odot}$; and the dark matter halo has a scale radius of 23 kpc and a central density of $3.8 \times 10^{-25}$ g cm$^{-3}$.  The gas disk has about 10\% of the mass in the stellar disk, and radial and vertical scales of 7 kpc and 0.4 kpc, respectively.  

To identify gas that has been stripped from the galaxy we also follow a passive tracer that is initially set to 1.0 inside the galaxy and $10^{-10}$ outside.  Specifically, the passive tracer is assigned in the initial problem set-up so that all gas with densities higher than the ICM density is set to one.  As gas from the disk is mixed into the ICM this allows us to determine, for each cell, the fraction of gas that originated in the disk (which we call the tracer fraction).  In the following analysis, we will use a minimum tracer fraction of 25\% to find gas stripped from the galaxy (as in Tonnesen et al. 2011; Tonnesen \& Bryan 2012).

\begin{figure*}
\begin{center}
\includegraphics[scale=1.08,trim= 0mm 0mm 3mm 0mm, clip]{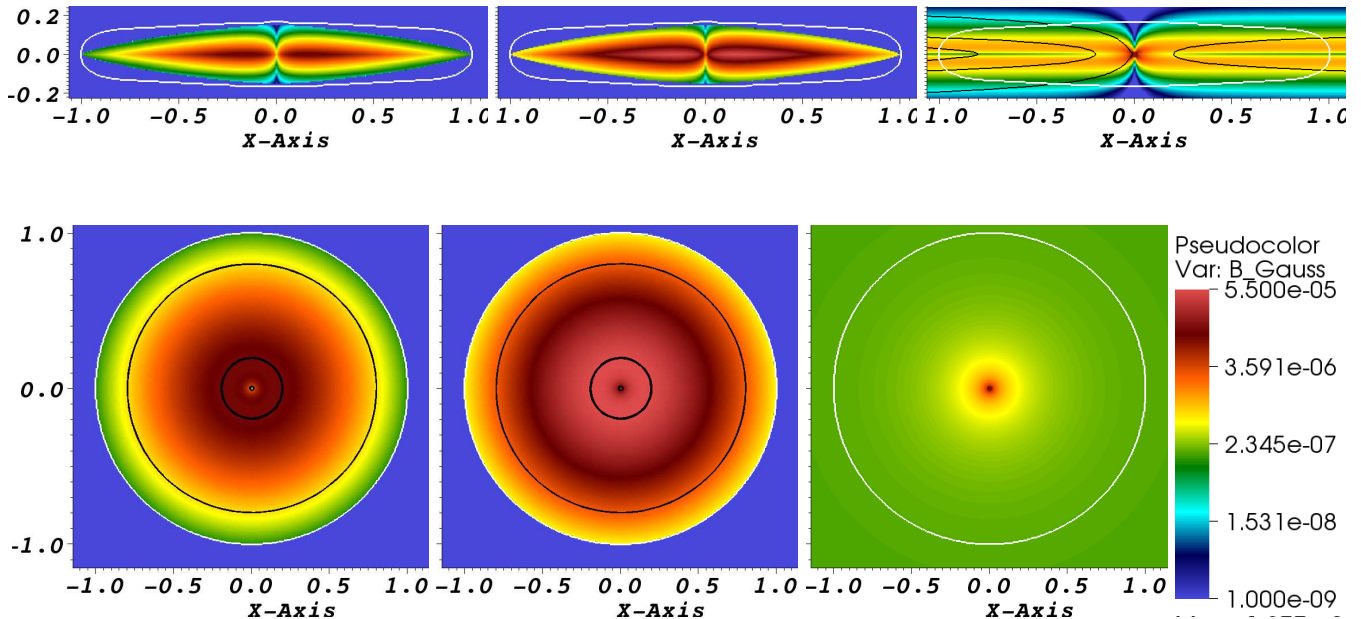}
\caption{Slices along the y- and z-axis for the initial conditions of the TORL, TORH, and DIP runs.  The color scale is the magnetic field strength, the black lines are streamlines of the magnetic field lines, and the white line shows the edge of the disk defined by where the density drops to that of the ICM.}\label{fig-bfield}
\end{center}
\end{figure*}

\subsubsection{The Galactic Magnetic Fields}\label{sec:magfield}

We analyze four simulations in this paper, in which we vary the initial galactic magnetic field.  Our baseline is a purely hydrodynamic run, Hydro.   
We also examine two runs with toroidal magnetic fields of different initial strengths: TORL and TORH.  Finally, we have run a simulation with a dipole field that has been compressed in the z-direction and stretched along the disk plane: DIP.  In order to initialize a divergence-free magnetic field, we input the vector potential, and calculate the magnetic field within the run.  

In the TORL and TORH runs we forced the magnetic field to be zero outside of the disk by setting the vector potential to a constant at a threshold A$_z$.  We selected a vector potential such that our magnetic field was weak in the center of the galaxy where the velocity is changing rapidly, peaked a couple kpc from the center (well within the stripping radius), and then fell off gradually with increasing radius.  The vector potential follows equations 1-4 in the disk.

\begin{equation}
A_x = A_y = 0
\end{equation}
\begin{multline}
A_z = \sqrt{a_{zf}} e^{(-6R_{cyl})} \times\\
\frac{( -6sin(2.5R_{cyl}) - 2.5cos(2.5R_{cyl}))}{(6^2 + 2.5^2)}
\end{multline}
\begin{equation}
a_{zf} = a_o  (-|z| + 1)^{80}
\end{equation}
\begin{equation}
a_{o(TORL)} = 1000, a_{o(TORH)} = 4000
\end{equation}

In the DIP run we began with a dipole vector potential, then forced it to decrease more rapidly in the z-direction and more slowly in the R$_{cyl}$-direction than a true dipole magnetic field.  In order to allow for closed magnetic field lines we did not set the vector potential to a constant outside of the disk.  Instead, in order to slow the growth of the magnetic field at the disk edges, we allow the surrounding ICM to rotate with the disk out to 2.4 R$_{disk}$ ($\sim$62 kpc).  The vector potential is described by equations 5-7.

\begin{equation}
A_x = \frac{-a_o y  R^2_{cyl}}{((z^2+0.01)^3  r^3_{sph})}\\
\end{equation}
\begin{equation}
A_y = \frac{a_o  x  R^2_{cyl}}{((z^2+0.01)^3  r^3_{sph})}\\
\end{equation}
\begin{equation}
A_z = 0, a_o = 3.0\times10^{-8}
\end{equation}

In Figure \ref{fig-bfield}, we show slices of the magnetic field magnitude including representative streamlines through the y=0 and z=0 planes of the galaxies.  We see that the central regions of the galaxies tend to have high magnetic field strengths in comparison to the 2-7 $\mu$G measured in the Milky Way (e.g. Men et al. 2008; Mao et al. 2012).  However, the strength of magnetic fields in nearby spiral gas-rich galaxies with high star formation rates range from 20-30 $\mu$G in the spiral arms, and starburst galaxies can have total fields of 50-100 $\mu$G (see review by Beck 2009).  Therefore, while our simulated magnetic fields may be somewhat strong for Milky Way measurements, they are not unphysical.  We also note that by a radius of 5 kpc the peak magnetic field magnitude in DIP is at or below 3 $\mu$G, and by 15 kpc the magnetic field in TORL peaks at $\sim$3 $\mu$G.  While the magnetic field in TORH is stronger than we would expect for a Milky Way type galaxy, it will allow us to place upper limits on the importance of the magnetic field strength to our results.  We consider the DIP and TORH runs to be our strong magnetic field cases because both have $\mu$G fields extending to the edge of the disk (the top panels of Figure \ref{fig-bfield}).

Our magnetic field morphologies are simplified in order to slow the growth of instabilities, which we discuss below.  For example, we do not include small-scale random fields (discussions in, e.g. Beuermann et al. 1985; Sun et al. 2008).  We also either cut-off (TORL and TORH) or severely lower (DIP) the strength of our magnetic fields at the disk edge--to maintain a lower magnetic pressure than thermal pressure at the disk edge in order to reduce disk expansion and to reduce magnetic field growth from the velocity differential between the disk and the ICM.  Observed magnetic fields tend to have a planar component that follows the spiral arms, with the field often strongest in the regions between the optical arms (Beck 2009 and references therein).  There have also been observations of an X-shaped vertical field in the halos of disk galaxies (Beck 2009 and references therein).  As we do not include radiative cooling or spiral density waves, it is sensible to begin with straightforward disk fields.  We choose the field morphologies described above in order to use reproducible, divergence-free fields that vary slowly due to instabilities.       
\begin{figure}
\begin{center}
\includegraphics[scale=1.0,trim= 0mm 0mm 111mm 0mm, clip,angle=270]{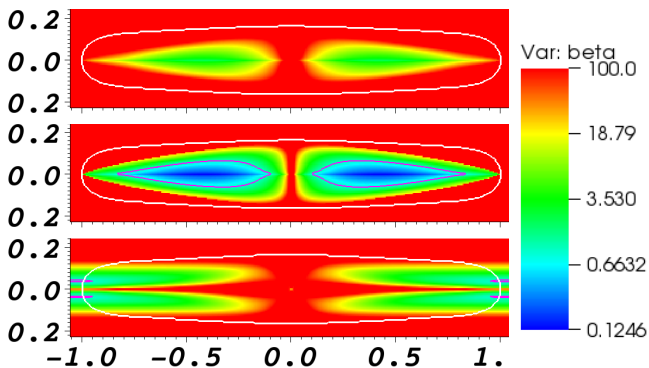}\\
\caption{Slices along the y-axis for the initial conditions of the TORL, TORH, and DIP runs from top to bottom.  The color scale is $\beta$$=$P$_{Thermal}$/(B$^2$/8$\pi$), The pink contour is $\beta$$=$1 (for TORH and DIP, the two runs with $\beta$$\ge$1), and the white line shows the edge of the disk defined by where the density drops to that of the ICM.}\label{fig-beta}
\end{center}
\end{figure}

\begin{figure*}
\begin{center}
\includegraphics[scale=1.07,trim= 0mm 0mm 5mm 22mm, clip]{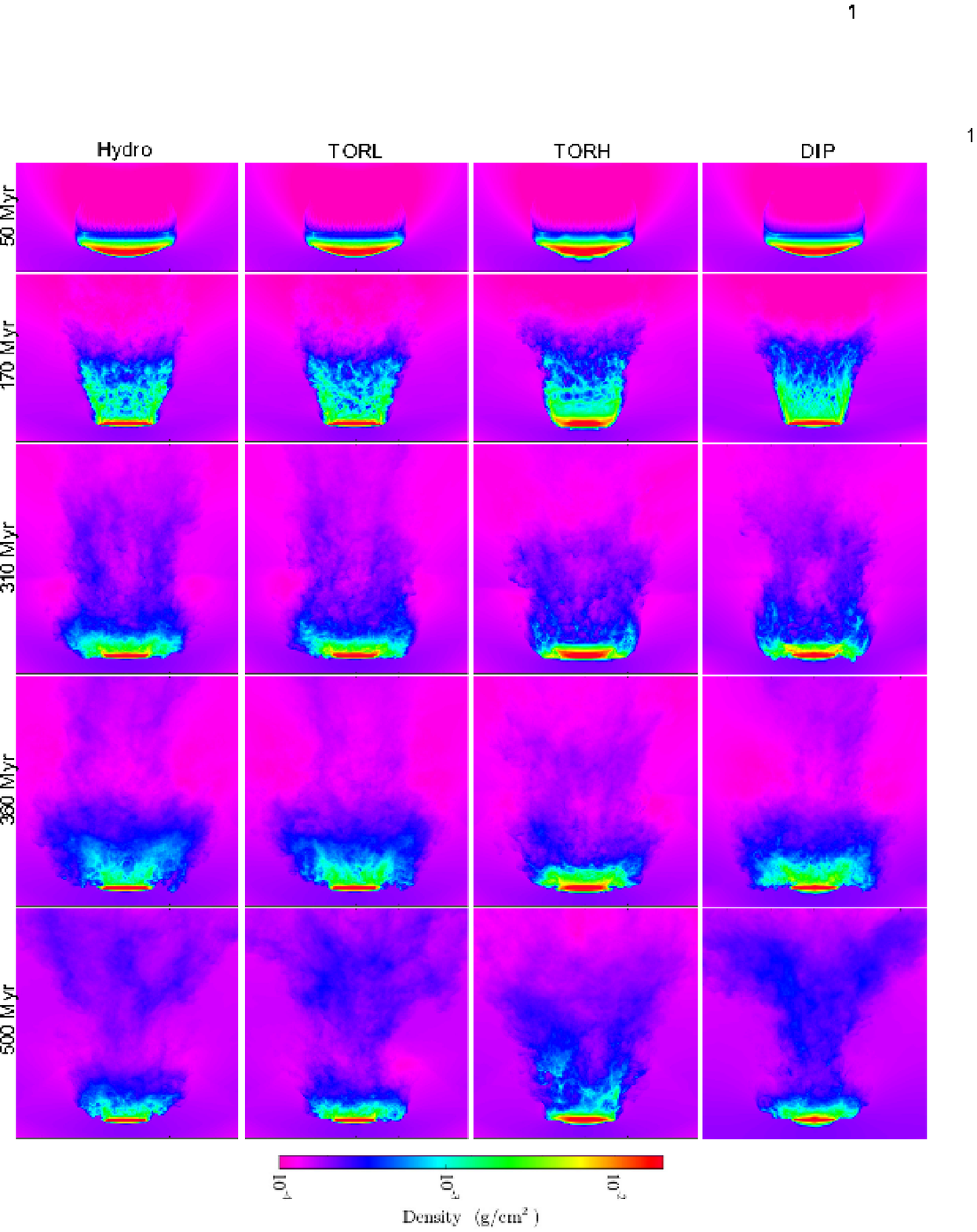}
\caption{Density projections for all four runs.  From left to right the columns show the Hydro, TORL, TORH, and DIP runs.   Each row shows a different time after the wind has hit the galaxy:  50 Myr, 170 Myr, 310 Myr, 360 Myr, 500 Myr. }\label{fig-projections}
\end{center}
\end{figure*}

In Figure \ref{fig-beta} we show the initial $\beta$$=$P$_{\mathrm{Therm}}$/(B$^2$/8$\pi$) values in our disks.  We have used TORH to set the color scale, as it has the lowest minimum $\beta$ of 0.1246.  The minimum $\beta$ values for TORL and DIP are 1.99 and 0.97, respectively.  We highlight where $\beta$$=$1 using a pink contour.  The added magnetic pressure does not have a strong effect on the disk in either the TORL or DIP runs, as it is generally well below that of the thermal pressure.  However, in the TORH run the magnetic pressure along the central plane of the disk from a radius of $\sim$2 -15 kpc is larger than the gas pressure by up to a factor of eight, so the disk quickly puffs up (recall that the gas thermal pressure is initially set to balance the gravitational potential).  In fact, the gas disk expands to more than 5 kpc from the disk plane by $t$=25 Myr, before the ICM wind hits the disk.  This is why the disk gas measured within 5 kpc of the disk plane begins decreasing the earliest in the TORH run, as seen in Figure \ref{fig-diskmass}.  By later times ($\sim$150 Myr after the wind has hit the disk), this difference does not affect how much gas remains in the disk.  The extra magnetic pressure in the TORH run also results in the thicker disk in the projections in Figure \ref{fig-projections} and in the higher velocities of high-density disk gas 50 Myr after the wind has hit the disk, in Figure \ref{fig-diskrhovz}.

Our initial conditions are not in magnetostatic equilibrium, so we see evolution in the magnetic field strength and structure, and in the gas distribution of the disk.  
While our DIP run begins with a magnetic field with only cylindrical radial and vertical components, due to differential rotation in the disk the azimuthal component grows with time.  This increases the magnetic pressure with time, and can be seen most clearly by the increase in the maximum pressure of the DIP run with time, and the increasing z-velocity of central, high-density, disk gas (Figures \ref{fig-diskrhovz} \& \ref{fig-diskpvz}).  The effect is strongest in the central regions of the disk, where it will have little effect on the stripping profile of the galaxy.  However, the evolving magnetic field does make us unable to use the DIP run when we discuss the effect of compression by the ICM wind on the magnetic field strength in the central regions of the disk (Section \ref{sec:obs}).

These disks are unstable to the magnetorotational instability (MRI) (Balbus \& Hawley 1991) for most of their radius, beyond $\sim$4 kpc, where the angular velocity begins decreasing outwards (Hawley \& Balbus 1999a,b).  The local growth rate of the MRI is proportional to the local orbital frequency (Hawley 2001), so the instability grows the most quickly near the center of the disk.  At 4 kpc, the orbital period is 120 Myr, and at the edge of the disk the orbital period is 800 Myr, so while the inside of the disk will have more gas mass and higher densities (seen in the increase in high density gas in the late panels in the DIP run in Figure \ref{fig-diskrhovz}), there will be very little expansion of the outer edge of the disk.  Even at 16 kpc, the initial stripping radius of the disk, the orbital period is 490 Myr, so disks do not expand much in our simulation.  If a disk remained on a relatively circular orbit for a long period of time, the MRI might result in more gas removal at late times because of disk expansion.  
 
\subsection{The ICM}\label{sec:ICM}

All four of the simulations we discuss in this paper initially embed a galaxy in a static, high-pressure medium with $\rho=$ 9.152 $\times$ 10$^{-29}$ g cm$^{-3}$ and $T = $ 4.15 $\times$ 10$^6$ K.  The boundary conditions generate a constant unmagnetized ICM inflow along the inner z-axis, which is always face-on to the galaxy.  The wind parameters are $P_{\rm ram} = \rho v^2_{\rm ICM} = 6.4 \times 10^{-12}$ dynes cm$^{-2}$, and $v_{ICM} = 1413$ km s$^{-1}$.  The ICM wind has a T = 4 $\times$ 10$^7$ K and $\rho$ = 3.2 $\times$ 10$^{-28}$ g cm$^{-3}$, and therefore P = 1.76 $\times$ 10$^{-12}$ dynes cm$^{-2}$.  These are the same ICM parameters as in Tonnesen \& Bryan (2009; 2010; 2012).

\begin{figure*}
\begin{center}
\includegraphics[scale=1.05,trim= 0mm 0mm 6mm 0mm, clip]{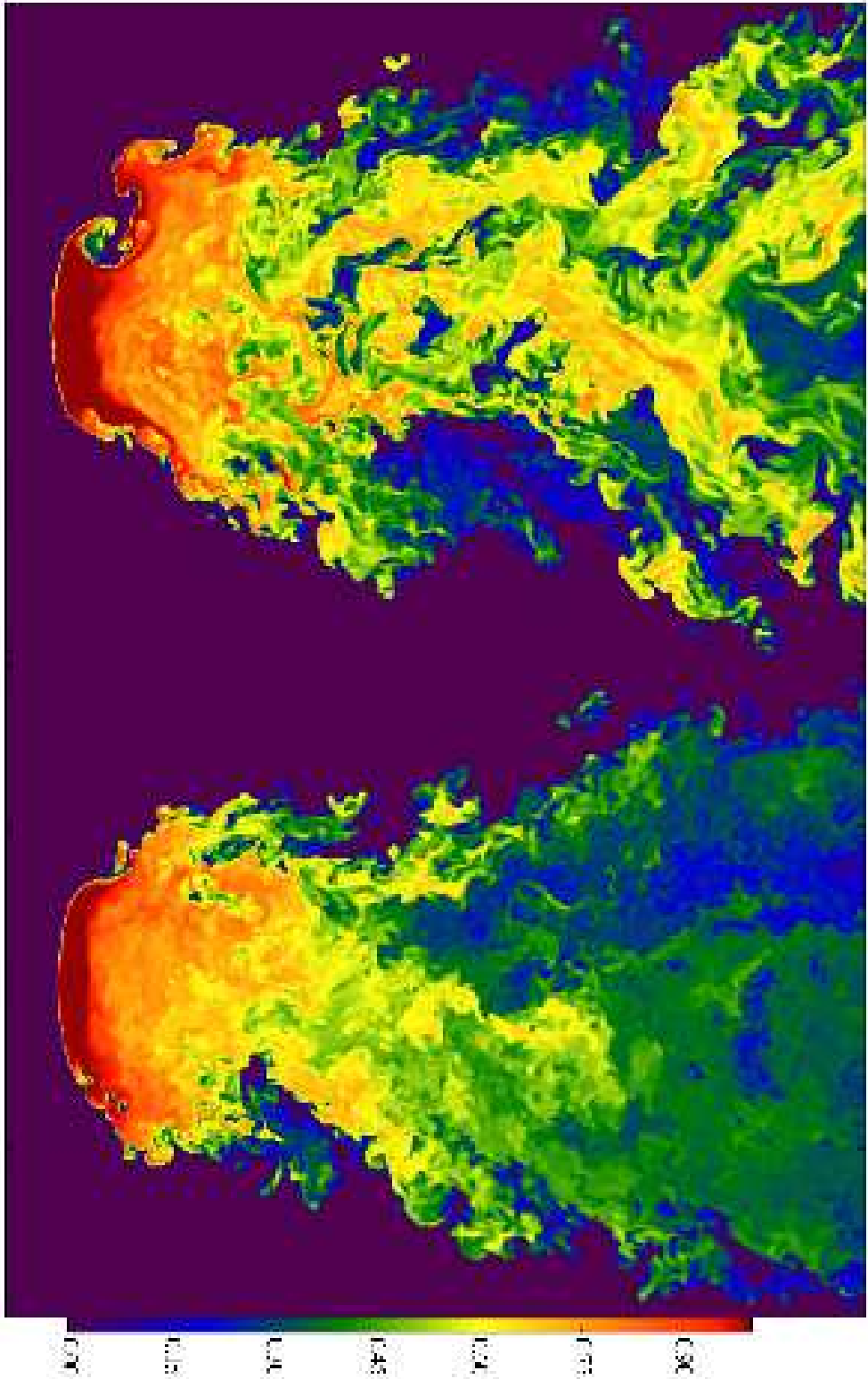}
\caption{A slice of the fraction of gas that originated in the disk in our TORH (upper panel) and Hydro (lower panel) simulations, 750 Myr after the ICM wind has hit the disk.  The color scale shows the fraction of gas in any cell that originated from the disk.  The range of mixing levels is broad at most heights above the disk, ranging from $\sim$0.2 to more than 0.5.  Much more nearly unmixed gas (more than 75\% of the gas in the cell originated in the disk) survives to the edge of our box in the TORH run, $\sim$100 kpc above the disk plane.  The slice is about 81$\times$107 kpc}\label{fig-mixing}
\end{center}
\end{figure*}

\section{The Effects of a Galactic Magnetic Field}\label{sec:results}

In Figure \ref{fig-mixing}, we show slices at y=0 of the fraction of gas that originated in the disk in our TORH (upper panel) and Hydro (lower panel) simulations, 750 Myr after the ICM wind has hit the disk.  Clearly we follow a range of mixing levels throughout the tails within our simulated region.  The structure in the tail indicates that disordered motions are mixing the gas in both the hydrodynamical and MHD runs.  In the TORH run, the eddy structures are larger than in the Hydro run, and much more nearly unmixed gas (more than 75\% of the gas in the cell originated in the disk) survives to the edge of our simulated region, $\sim$100 kpc above the disk plane.

While Figure \ref{fig-mixing} is useful to see the tail structure in detail, density projections can be more directly compared to observations.  Therefore, for reference in the following discussions, we include 5 density projections for each of our runs in Figure \ref{fig-projections}.  Each column is a different simulation and each row steps through time, as noted in the caption and in Figures \ref{fig-diskmass} and \ref{fig-diskrad}.  The first row, 50 Myr after the wind has hit the disk, shows that there is little initial difference between the Hydro and TOR (L \& H) cases, but the tail is already more extended in the DIP run.  This behavior is caused by the magnetic field threading radially through the disk into the surrounding low-density gas in the DIP run.  As the low density gas is swept up by the wind, magnetic tension will increase the total pressure removing the disk gas.  170 Myr after the wind has hit the disk (second row), the Hydro and TORL simulations are very similar, but the stronger magnetic field in the TORH disk results in less flaring through most of the tail.  The DIP run also has less flaring in the tail, and an organized ``streak"-pattern in the stripped gas that can be seen criss-crossing in the tail due to the radial component of the magnetic field.  At 310 Myr (third row), the strong magnetic field cases (TORH and DIP) have shorter dark blue tails (mid-range densities) than the Hydro and TORL runs, and this continues to be true 360 Myr after the wind has hit the disk.  However, at 360 Myr after the wind has hit the disk, the surface area of higher density gas (yellow and green) above the disk is larger in the TORH and DIP runs than in the Hydro and TORL runs.  500 Myr after the wind has hit the disk (bottom row), higher density gas is clearly seen farther above the disk in the TORH run (light blue and green), and in a careful examination, light blue higher column density regions can be seen in the DIP run even farther along the tail than in the TORH run, out to $\ge$65 kpc above the disk.                  

\subsection{The Gas Disk}\label{sec:gasdisk}

In this section we discuss the differences in the gas residing in the disk in our four simulations.  Disk gas is defined as gas with a tracer fraction greater than 0.6 and within a cylinder with 28.6 kpc radius and 10 kpc height ($\pm$5 kpc from the disk plane).  Our choice of limiting tracer fraction does not have any qualitative impact on our results.  The gas disk mass using a tracer fraction of 0.25 is always less than 2\% different than 0.9.  We see in Figure \ref{fig-mixing} that most of the gas near the disk has a tracer fraction more than 0.9, and the transition from 0.9 to 0 occurs over a short distance at the disk edges.  Gas that falls back tends to have a lower tracer fraction (be more mixed with the ICM), but the center of infalling clumps have high tracer fractions so a limiting value of more than 0.75 would need to be chosen to have much impact on the measured radius.  The phase plots do change when we dramatically vary the minimum tracer fraction used to define disk gas--there is more low-density gas when we use 0.25, and almost no gas with densities below 2 $\times$10$^{-27}$ g cm$^{-3}$ when we choose 0.9.  However, the red and orange contours remain unchanged and the comparisons between the different runs remain unchanged.  

\subsubsection{The Disk Mass}

\begin{figure}[!h]
\begin{center}
\includegraphics[scale=0.74,trim= 0mm 0mm 0mm 0mm, clip]{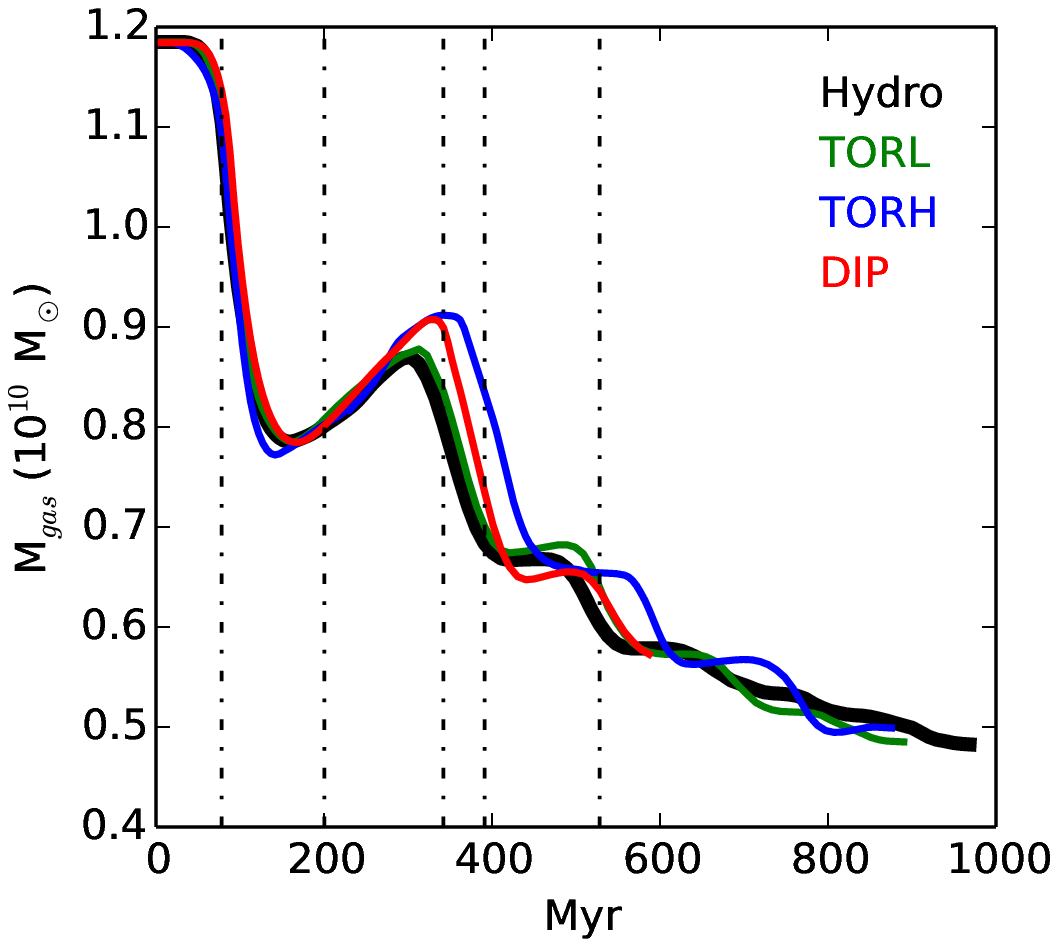}
\caption{\scriptsize{The amount of gas in the disk as as function of time.  The disk gas is defined as gas with a tracer fraction greater than 0.6 and within a cylinder with radius 28.6 and height 10 kpc ($\pm$ 5 kpc from the disk plane).  The wind hits the disks at t$\sim$30 Myr, so 50 Myr after the wind has hit the disk is at 80 Myr on the x-axis of this figure.  The vertical dashed lines denote the times at which we show density slices in Figure \ref{fig-projections}.  Although the initial stripping seems to depend on the magnetic field strength, by about 310 Myr after the wind has hit the disk (the third vertical line), the amount of gas remaining in the disk is very similar in all four runs.   }}\label{fig-diskmass}
\end{center}
\end{figure}

In Figure \ref{fig-diskmass}, we plot the amount of gas in the disk as a function of time for all four runs.  
The wind hits at t $\sim$ 30 Myr (t $\sim$ 35 Myr in the DIP run).  The vertical lines denote the times at which density projections are shown in Figure \ref{fig-projections}, and a selection of these times will be used in the phase plots throughout the paper.  

We first highlight the similarities between the four cases.  In all four cases there is an initial stripping event lasting between 110-170 Myr with the fastest stripping rate in the simulation.  All four simulations then undergo fallback onto the disk.  As we have discussed in earlier work (Tonnesen \& Bryan 2009, 2010; also Schulz \& Struck 2001), fallback occurs when stripped gas that is still gravitationally bound to the galaxy moves into the protected lee of the disk and is no longer pushed by the wind. 
At 300 Myr after the wind has hit the disk the stripping rate begins to slow and to have less dramatic fallback episodes.  Also, after 300 Myr of stripping there is very little difference in the amount of disk gas in the four runs, with the largest difference between runs being less than 10\% of the total gas disk mass.  At later times gas is removed by continuous stripping processes, e.g. Kelvin-Helmholtz instabilities or viscous stripping, as discussed in Tonnesen \& Bryan (2009).  Including either a toroidal (TORL or TORH) or poloidal (DIP) magnetic field has little effect on the rate of this instability-driven gas removal.  
   
The main difference in these runs is that more gas falls back in the strong magnetic field cases (TORH and DIP) than in the weak field cases (Hydro and TORL).  We suspect that this is because the stronger magnetic tension initially confines the stripped gas to a more collimated tail behind the disk (as seen by the narrower tails in the second column of Figure \ref{fig-projections}).  This allows more gas to be in the shadow of the disk for added fallback.

There is an initial difference in the TORH run due to the initial conditions of the field, as discussed in Section \ref{sec:magfield}.  The TORH disk seems to lose mass the most quickly, even though the magnetic field does not extend beyond the disk and the field is toroidal, so magnetic tension should not initially play a role in gas stripping.  In agreement with this expectation, the dark blue column density gas in the first TORH and Hydro projections looks nearly identical (Figure \ref{fig-projections}).  However, TORH has a strong magnetic field, such that the magnetic pressure in the inner $\sim$15 kpc puffs up the disk so that some gas is more than 5 kpc from the disk plane, leading to a decrease in the disk gas measurement (which is defined to only include gas within 5 kpc of the disk plane) before the ICM wind can strip the disk.  This can also be seen in Figure \ref{fig-projections} when comparing the light blue and green gas above the central region of the disk in TORH to any other simulation.  At later times (beyond 300 Myr) this lower-density gas is stripped from all four galaxies, so the TORH run evolves similarly to the other runs.  Although the highest density gas (red in the projection) also has a larger scale height in the TORH run, it remains within 5 kpc of the disk plane, and so does not affect this comparison between the runs.

\subsubsection{The Disk Radius}

\begin{figure}[!h]
\begin{center}
\includegraphics[scale=0.74,trim=0mm 0mm 0mm 0mm, clip]{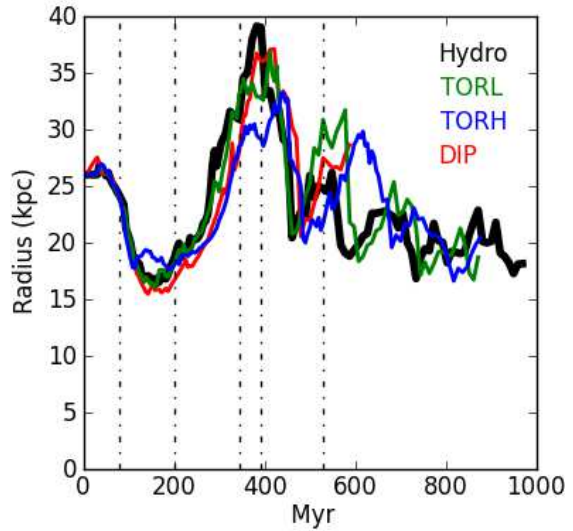}

\caption{\scriptsize{The disk radius as a function of time, calculated as the largest radius at which gas with a tracer fraction greater than 0.6 resides within 5 kpc of the disk plane.  The minimum radius reached in the initial stripping in all four cases is very similar.  The bulk of the gas near the disk plane ($|z|$ $<$ 5 kpc) stays within the initial stripping radius, between 15-19 kpc for all the runs.  The peaks are from material that has been stripped falling back towards the disk plane.  The peak radius, reached in the initial fallback period, is largest in the Hydro case, as might be expected with no magnetic field confinement. } } \label{fig-diskrad}
\end{center}
\end{figure}

The disk radius is calculated as the largest radius at which gas with a tracer fraction greater than 0.6 resides within 5 kpc of the disk plane.  As with the gas mass shown in Figure \ref{fig-diskmass}, the stripping and fallback cycle leads to variations in the gas radius.  Through comparison with Figure \ref{fig-diskmass} we find that fallback coincides with an increase in gas mass at large galactic radius.  Comparing with the final three rows in Figure \ref{fig-projections}, we see that the fallback at large radius is clumpy and asymmetric.  We also examine disk symmetry by finding the maximum radius in four quadrants of each disk.  For the first 250 Myr of stripping, before the first fallback peak in Figure \ref{fig-diskmass}, the four quadrants in each disk have visually identical (to within the width of the line) maximum radii.  After this time in the simulations, there is a radial variation of about 2 kpc between the largest and smallest quadrants at any given time, although the variation can be up to $\sim$10 kpc for short times (tens of Myr) near the peak in radius at $\sim$400 Myr.  In this face-on ICM-ISM interaction, fallback, not stripping, drives asymmetry.

The initial minimum radius is reached at nearly the same time as the first minimum in disk gas mass from the initial stripping, but the peak in radius is somewhat later than the fallback peak in gas mass.  This may be because material falls towards the disk then splashes outward along the disk plane from the added pressure.  Although the fallback looks dramatic in Figure \ref{fig-diskrad}, the bulk of the disk gas mass remains within the initial stripping radius, which can be seen in the constant size of the red regions in the projection plots in Figure \ref{fig-projections}.  The radial variations with time continue throughout all of our runs, although they seem to be settling towards the initial stripping radius.  As discussed above, the gas removal at later times is due to Kelvin-Helmholtz instabilities or viscous stripping.  As the ICM wind flows around the disk there is a component of the wind moving along the disk plane, so these processes may draw disk material to larger radii before the wind accelerates the gas from the disk plane, resulting in the small variations seen in the disk radius on short timescales.

There is very little difference in all four runs, but we see the effect of the strong magnetic fields in TORH by the smaller maximum radius reached by gas falling back onto the disk.  As discussed with regards to the the TORH run in the second row of Figure \ref{fig-projections}, the magnetic tension inhibits the radial expansion of the stripped gas.

\subsubsection{A Closer Look at the Disk Gas}\label{sec:diskrho}

\begin{figure*}
\begin{center}

\includegraphics[scale=1.07,trim= 0mm 0mm 7mm 0mm, clip]{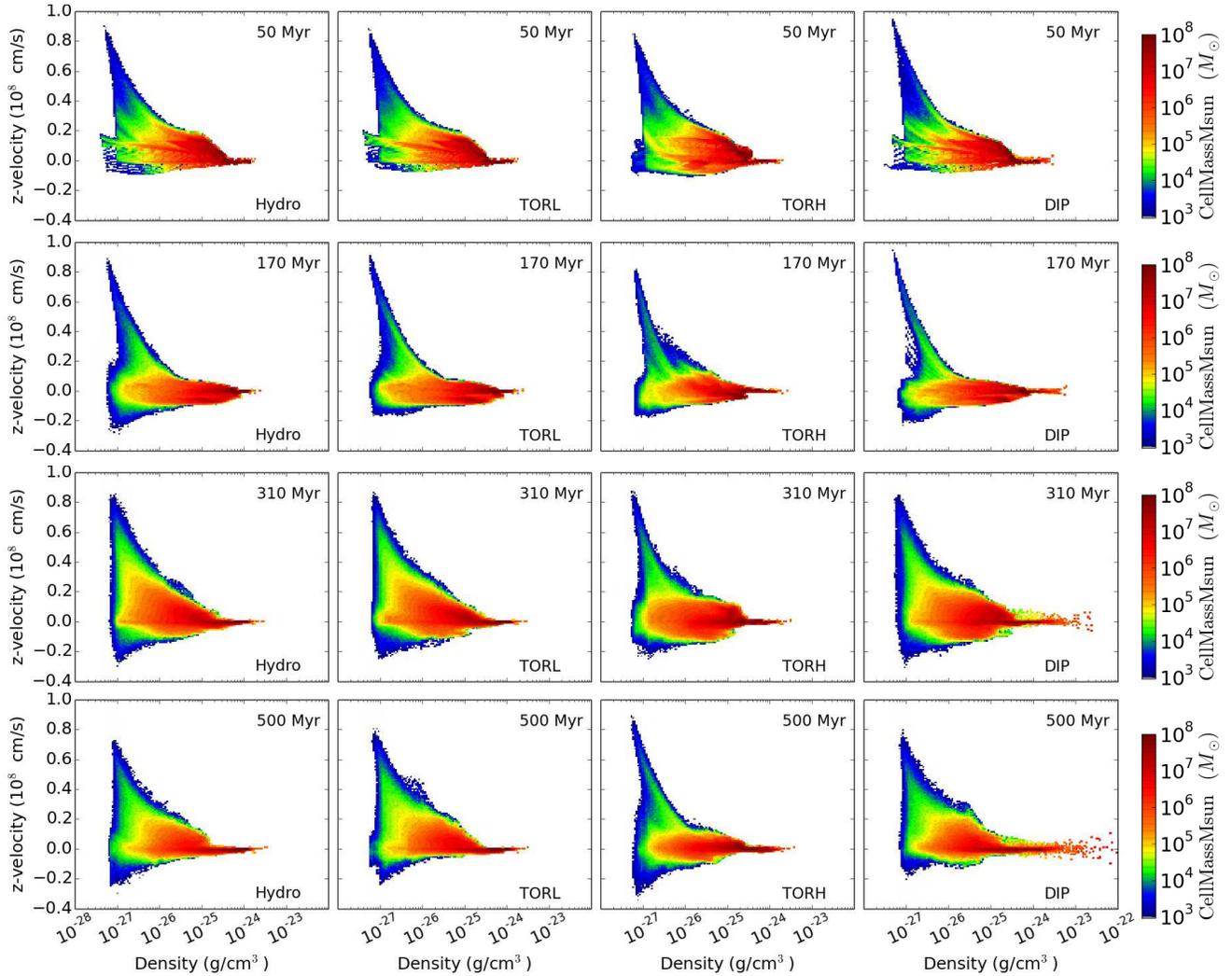}
\caption{\scriptsize{Mass contours of disk gas as a function of density and velocity in the wind direction (z-velocity).  Disk gas is gas with a tracer fraction greater than 0.6 and within a cylinder of radius 31 kpc and height 10 kpc ($\pm$ 5 kpc from the galaxy plane).  The columns, from left to right, are from the Hydro, TORL, TORH, and DIP runs.  Each row is a different time after the wind has hit:  50 Myr, 170 Myr, 310 Myr, 500 Myr.  While all four runs show strong similarities, the differences indicate the impact of including magnetic fields.  See the discussion in Section \ref{sec:diskrho}.}}\label{fig-diskrhovz}
\end{center}
\end{figure*}

In Figure \ref{fig-diskrhovz} we show disk gas mass as a function of density and z-velocity.  In this figure, disk gas is gas with a tracer fraction greater than 0.6 and within a cylinder of radius 31 kpc and height 10 kpc ($\pm$ 5 kpc from the galaxy plane). 
  
At 50 Myr there is no significant difference between the Hydro and TORL runs, but we see more clearly the differences in the TORH and DIP runs that we have discussed with regards to Figure \ref{fig-projections}.  The blue contours in TORH are very similar to those in TORL and Hydro, indicating that gas at the edge of the disk is being stripped at the same rate in all three runs.  However, there is more gas with negative velocities, particularly at 10$^{-26}$$<$$\rho$$<$10$^{-25}$, and there is more high-density gas with large positive velocities.  This indicates that magnetic pressure is driving expansion of dense (and therefore central) disk gas.  Low-density gas ($\rho$$<$10$^{-27}$) in the DIP run has reached a larger maximum velocity by 50 Myr, indicating that the magnetic field threading radially through the disk and into the surrounding medium is increasing the acceleration of the lower-density stripped gas.  

At 170 Myr, the DIP run still has gas being removed from the disk at the highest velocities.  Here it is clear that the gas in the TORH run is being accelerated the most slowly, possibly due to pressure support from the magnetic field.  There is also less fallback in the TORH run, shown as less gas with negative velocities, particularly at $\rho$$<$10$^{-26}$.  Indeed, including any magnetic field in the disk results in slower fallback at this early time, as the Hydro run has the largest negative velocities.    

At 310 Myr, the Hydro and TORL runs remain very similar.  At densities below 10$^{-25}$, the blue and green contours of DIP also look quite similar to the Hydro run, indicating that magnetic tension is no longer accelerating stripping in DIP.  In fact, at this later time when continuous stripping processes dominate gas removal, a comparison of the yellow and orange contours indicates that the bulk of the gas with a density of a few 10$^{-27}$ is being accelerated more slowly in the DIP run than in the Hydro and TORL runs.  However, the gas at $\rho$$\ge$10$^{-25}$ g cm$^{-3}$ has a broader z-velocity distribution in the DIP run than in the Hydro and TORL runs.  The TORH run has similar differences from the Hydro and TORL runs.  As in the DIP run, the bulk of the gas at $\rho$$<$10$^{-25}$ g cm$^{-3}$ is accelerated more slowly by the wind, but this difference is more pronounced when comparing TORH and Hydro (the difference is seen at all contour levels).  Also, the gas at $\rho$$\sim$10$^{-25}$ g cm$^{-3}$ has a broader z-velocity distribution in the TORH run than in the Hydro or TORL runs.  The dense gas in the TORH and DIP runs is expanding due to pressure from the galactic magnetic fields, the same fields that may be inhibiting the removal of low-density material by the ICM wind.  

Finally, at 500 Myr, the lower-density gas ($\rho$$<$10$^{-25}$ g cm$^{-3}$) in the DIP run has a very similar velocity distribution to that in the Hydro and TORL runs, and indeed Figure \ref{fig-diskmass} indicates that the stripping rate at later times is very similar in all four runs.  At higher densities the velocity distribution of gas is broader because the differential rotation in the inner part of the disk has strengthened the magnetic field, and magnetic pressure drives some gas from the disk center.  In TORH, which has the strongest initial magnetic field throughout the disk, less gas is accelerated quickly from the disk, although clearly the difference is not large enough to differentiate the gross stripping rate in TORH from that in TORL and Hydro, as seen in Figure \ref{fig-diskmass}.  As in DIP, the stronger central magnetic field results in a broader velocity distribution in the center of the disk, but this gas is too tightly bound to be stripped.

\begin{figure*}
\begin{center}
\includegraphics[scale=1.07,trim= 0mm 0mm 7mm 0mm, clip]{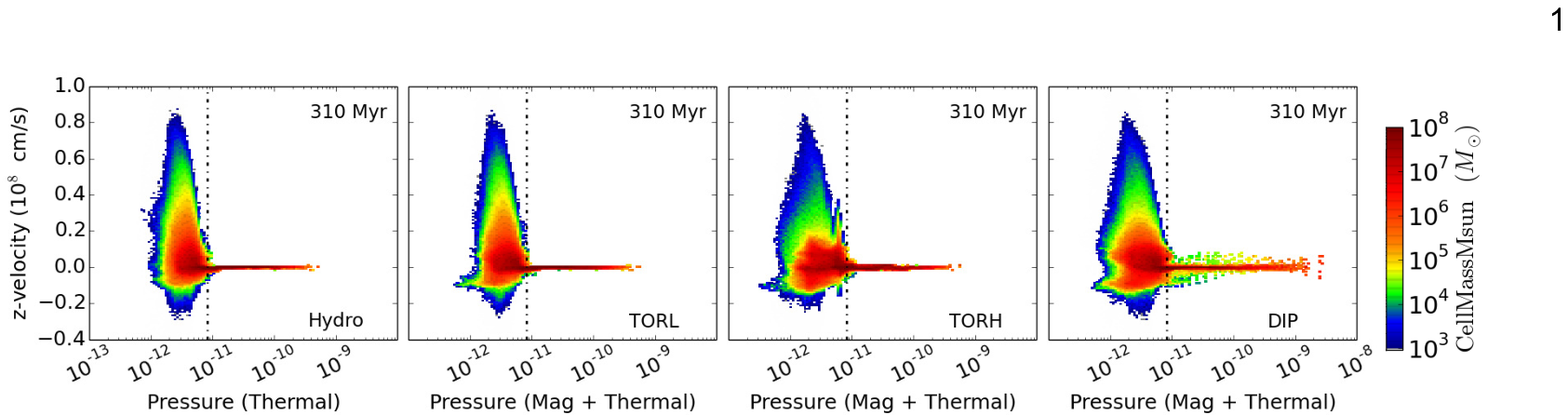}

\caption{\scriptsize{Mass contours of disk gas as a function of pressure and velocity in the wind direction (z-velocity) 310 Myr after the wind has hit the disk.  Disk gas is gas with a tracer fraction greater than 0.6 and within a cylinder of radius 31 kpc and height 10 kpc ($\pm$ 5 kpc from the galaxy plane).  The magnetic pressure has extended the pressure range in these figures to both higher and lower total pressure than the Hydro run.  As in the figure above, there is very little difference in the gross characteristics of the velocity structure of the disk gas in the four runs.  Only gas with lower total pressure than the sum of the ram pressure and ICM thermal pressure (dash-dot vertical line) is stripped.}  We can see evidence of magnetic pressure driving disk expansion in the larger velocity range of gas with P $>$ 10$^{-11}$.  }\label{fig-diskpvz}
\end{center}
\end{figure*}

It may also be physically informative to consider the pressure of the disk gas.  In Figure \ref{fig-diskpvz} we show the mass contours of disk gas as a function of pressure and velocity in the wind direction (z-velocity) 310 Myr after the wind has hit the disk.  
We have drawn a vertical line on each plot at P$_{crit}$ = P$_{ram}$ + P$_{\mathrm{Thermal, ICM}}$ = 8.16$\times$10$^{-12}$ dyne cm$^{-2}$.  Gas with pressure above this critical pressure cannot be removed from the disk.  This is because in our initial conditions, we set the pressure gradient to balance the gravitational potential in the disk z-direction, which links the disk pressure to the gravitational restoring force that binds that gas to the galaxy.  In any line through the disk along the wind direction, if the maximum gas pressure (thermal plus nonthermal) is greater than P$_{crit}$, gas cannot be removed by ram pressure.  Even continuous stripping mechanisms will lower the pressure of the gas through mixing as they remove gas from the disk.  It is not obvious that when radiative cooling and other sources of pressure are included, this result will hold, but if gas pressure can still be considered a proxy for the gravitational restoring force, then this would be a more direct way to determine whether gas can be stripped from a cluster galaxy.  Of course, determining the total pressure of galactic gas is difficult, because non-thermal pressure support can play a large role in balancing the galaxy's gravitational potential.

In summary, the stripping rate from the disk depends very little on either the morphology or strength of the magnetic field.  However, a poloidal field (DIP) that connects high-density to low-density gas can increase the early acceleration of stripped gas.  At later times when continuous stripping processes dominate gas removal, stronger magnetic fields seem to bind gas more tightly to the disk, resulting in the slower z-velocities of the bulk of the gas in the disk.  However, even global magnetic fields that are stronger than those found in the Milky Way do not have a drastic effect on the stripping amount or radius, nor do they strongly affect the density of the gas that can be removed from the disk.

\subsection{The Gas Tail}\label{sec:gastail}

In this section we focus on the gas tail, to determine if a magnetic field affects its structure and evolution.  Throughout this section, tail gas is defined as gas more than 10 kpc above the disk with a tracer fraction greater than 0.25.  

\subsubsection{The Tail Velocity Structure}\label{sec:tailzvz}

\begin{figure*}
\begin{center}
\includegraphics[scale=1.07,trim= 0mm 0mm 7mm 0mm, clip]{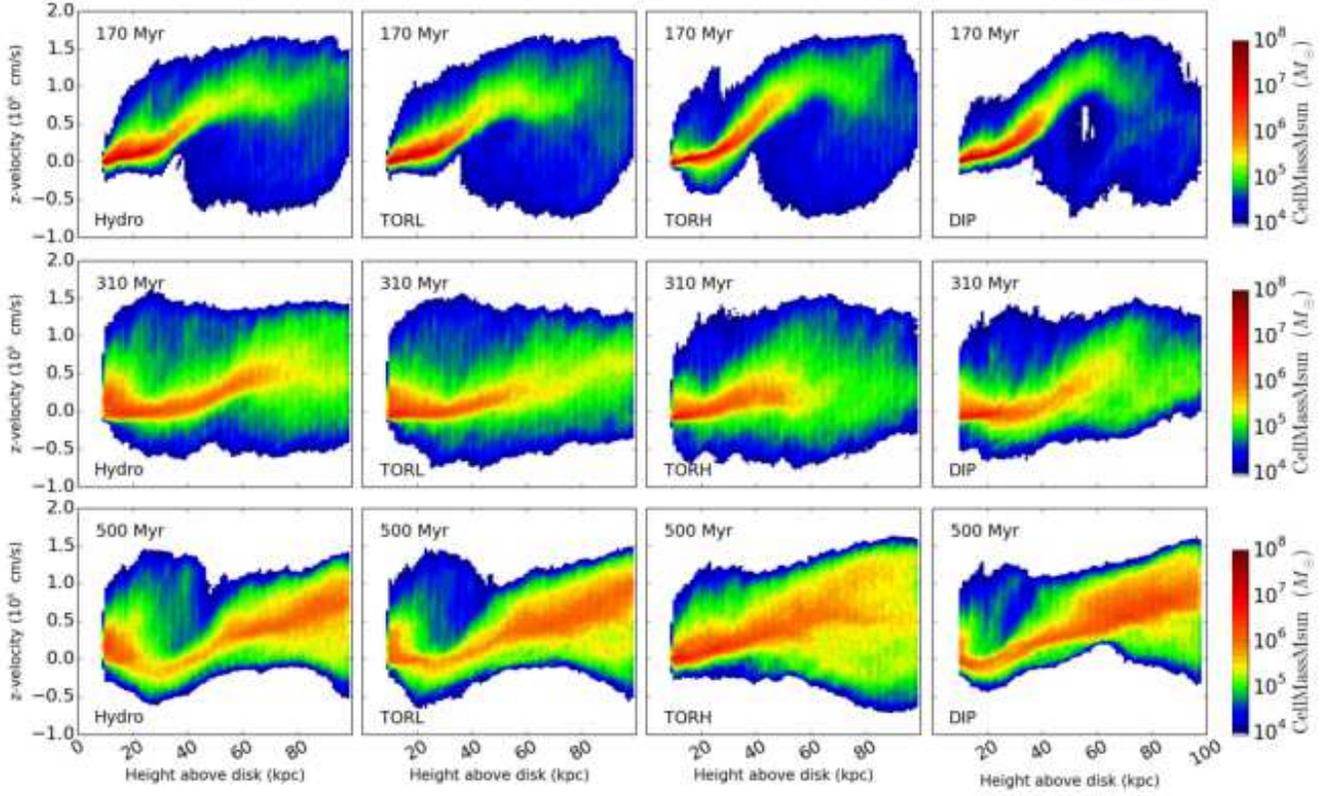}

\caption{\scriptsize{Mass contours of tail gas as a function of height above the disk and velocity in the wind direction (z-velocity).  Tail gas is gas more than 10 kpc above the disk with a tracer fraction greater than 0.25.  There is little difference between the four runs.  Hydro has more gas moving quickly close to the disk, but there is some indication that at larger distances the gas tends to move more quickly in the MHD runs.  See Section \ref{sec:tailzvz}.} }\label{fig-tailzvz}
\end{center}
\end{figure*}

In Figure \ref{fig-tailzvz}, we plot gas mass as a function of height above the disk and velocity in the wind direction (z-velocity).  We define tail gas as gas more than 10 kpc above the disk with a tracer fraction greater than 0.25.  As in the velocity structure of disk gas, there are not many dramatic differences in the velocity structure of the tail gas between the Hydro and MHD runs.  This is an important result because it means that estimates of tail velocities and lengths based on hydrodynamical simulation results do not have to be dramatically revised.  However, it is worthwhile to examine the velocity structure closely to determine any effects of magnetic fields.

170 Myr after the wind has hit the disk, all four runs are just past their first disk gas mass minimum (Figure \ref{fig-diskmass}).  In Figure \ref{fig-tailzvz}, TORH has more gas with negative velocities at $\sim$20 kpc above the disk, although DIP does not show more fallback than in the TORL and Hydro runs.  It is possible that most of the infalling gas in DIP is between 5-10 kpc above the disk at this time so is not included in either figure.  In Hydro the stripped gas at $\sim$20 kpc is moving with higher velocities than in the MHD runs, evidenced by both the orange and green contours.  However, more than 20 kpc above the disk, TORH has the most stripped gas farthest from the disk and moving at the highest velocities of any of the four runs, with DIP close behind.  This is clearly shown in the orange and yellow contours.  

310 Myr after the wind has hit the disk, the Hydro run has the longest orange contour, possibly indicating the survival of dense gas to $\sim$70 kpc above the disk.  Recall that the survival of dense gas to large distances in the Hydro run is also shown in the dark blue contours in the third row of Figure \ref{fig-projections}.  There is a slight indication that TORH and DIP have more fallback than Hydro and TORL, which agrees with the gas mass as a function of time in Figure \ref{fig-diskmass}--fallback is just finishing in TORH and DIP, while gas removal has already restarted in Hydro and TORL.  While the gas at $\sim$40 kpc above the disk is moving at similar velocities in all four runs, again there is more gas at high velocities within 20 kpc of the disk in the Hydro run.  

Finally, 500 Myr after the wind has hit the disk, the velocity structure in the TORH tail looks quite different than in the other three tails.  TORH has very little gas with negative velocities, while the majority of the gas from 20-40 kpc above the disk in the Hydro and TORL runs and from 15-25 kpc in the DIP run has negative velocities.  There is also less gas in the tail from 80-100 kpc above the disk in the TORH run than in any of the other three runs (also see Figure \ref{fig-projections}), and the gas at those large distance has the broadest z-velocity distribution of any of the runs.  As in the top panels of this figure, more than 20 kpc above the disk, gas tends to be moving more quickly from the galaxy in TORH and DIP than in Hydro and TORL.

In brief, including magnetic fields in the tail does not narrow the velocity width of the tail in the wind direction.  It also does not dramatically affect the bulk flow of the gas, although we note two points that are consistent in all three outputs we examine in detail:  first, within 20 kpc of the disk, the Hydro run has more gas accelerated to high velocities, and second, more than 20 kpc above the disk, the tail gas in the TORH and DIP runs tends to be moving away from the disk at velocities equal to or greater than the tail gas in the Hydro run.  

We posit that these differences are because near the disk the magnetic field slows the acceleration of gas, but farther from the galaxy the magnetic field in the tail allows larger coherent structures to survive that are then swept up by the ICM wind rather than mixed into the ambient ICM.  In Figure \ref{fig-mixing} this may be seen at 310 Myr in the green contours lying closer to the disk in TORH and DIP than in TORL and Hydro, and at 500 Myr in the dense gas (light blue) seen farther from the disk in the MHD runs than in Hydro.  In order to determine how this may affect observable tails, we need to include radiative cooling, which we will include in a future work.  

\subsubsection{Density and Temperature Structure of the Tail}\label{sec:rhoTtail}

\begin{figure*}
\begin{center}
\includegraphics[scale=1.1,trim= 0mm 0mm 10mm 0mm, clip]{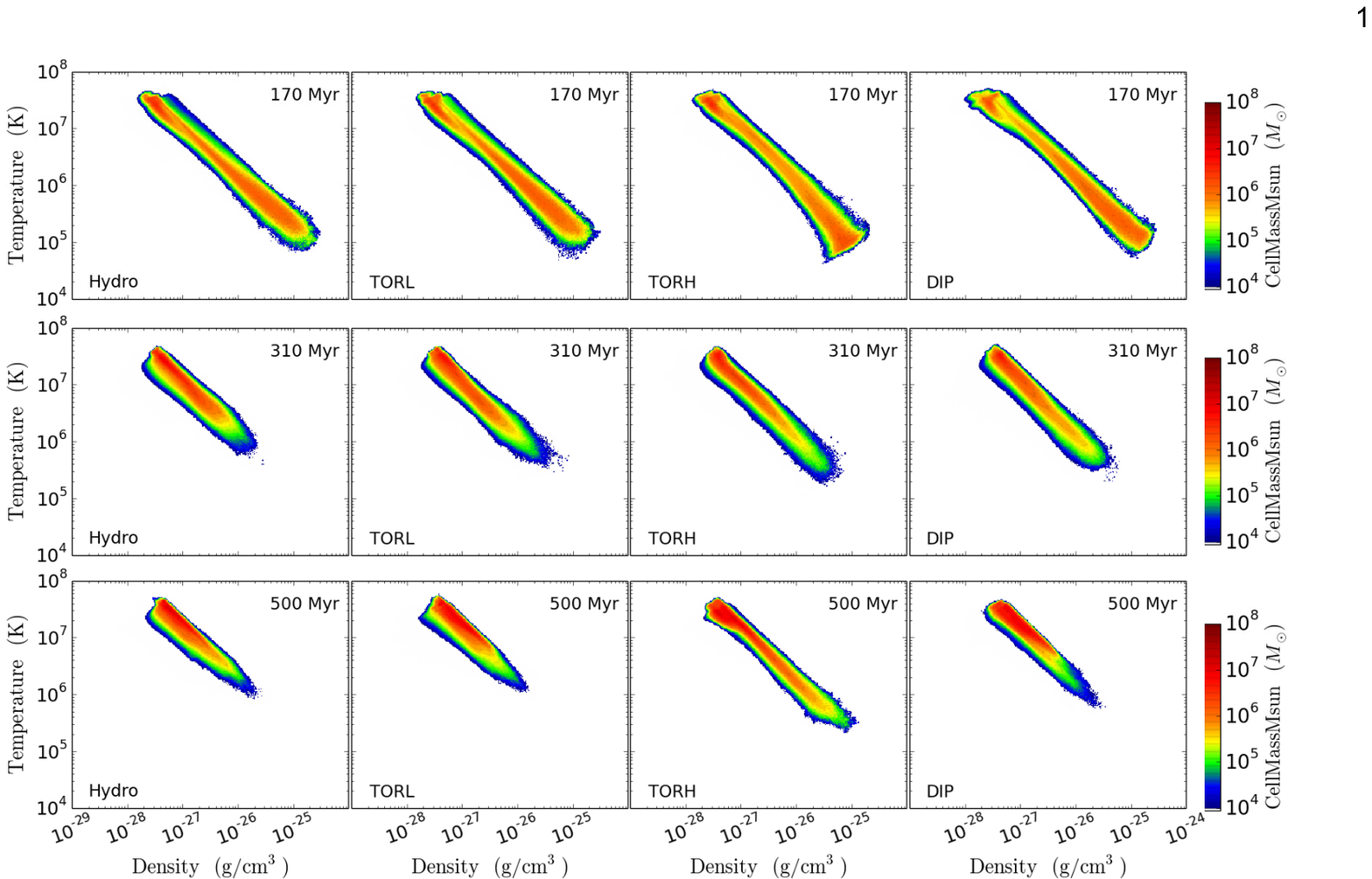}

\caption{\scriptsize{Mass contours of tail gas as a function of density and temperature.  Tail gas is defined as in Figure \ref{fig-tailzvz}. 
As time passes, there is more high-density gas in the tail in the magnetic field cases than in the Hydro run, indicating that mixing acts more slowly if magnetic fields thread the tail gas.  
} }\label{fig-tailrhoT}
\end{center}
\end{figure*}

In Figure \ref{fig-tailrhoT}, we plot the mass of gas in the stripped tail as a function of density and temperature.  As in Figure \ref{fig-tailzvz}, tail gas is defined as gas more than 10 kpc above the disk plane with a tracer fraction greater than 0.25.  

Our first row shows the density-temperature structure of gas 170 Myr after the wind has hit the galaxy.  At this time the density range of the gas is similar in all four runs, with the high-temperature, low-density tail gas matching the temperature and density of the ICM wind.  TORH and DIP have more low-temperature ($<$10$^5$ K) gas residing in the tail, although it is unclear whether in the Hydro and TORL runs this cold gas has been mixed into the ICM or simply has not yet reached 10 kpc above the disk.    
310 Myr after the wind has hit the galaxy, at the fallback peak in TORH and DIP in Figure \ref{fig-diskmass}, there is more cool ($<$10$^6$ K), higher-density ($\rho$$>$10$^{-26}$ g cm$^{-3}$) gas in all three MHD runs than in the Hydro run.  By 500 Myr after the wind has hit the galaxy, there is clearly much more cool ($<$10$^6$ K), higher density ($\rho$$>$10$^{-26}$ g cm$^{-3}$) gas in the tail in the TORH run than in any of the other three runs.  

We see that in all runs, at 170 Myr the density and temperature distribution of tail gas extends to higher densities and lower temperatures than at later times.  In order for gas to be removed from this density-temperature plane it must mix with the ICM, leave the box, or fall back to within 10 kpc of the disk.  Therefore to explain the difference between TORH and the other runs at 500 Myr the inclusion of the magnetic field must lower the mixing rate of stripped gas, lower the velocity of stripped gas, impede fallback, or drive some combination of these three possibilities.  

We show in Figure \ref{fig-mixing} that at very late times (750 Myr after the wind has hit the disk), more unmixed gas survives to large distances from the disk in TORH than in Hydro.  While this indicates that the magnetic field inhibits mixing, this is one slice from a complicated flow, as is clear from Figures \ref{fig-projections} and \ref{fig-tailzvz}.    
Examining Figure \ref{fig-tailzvz}, we find that the velocity of the stripped gas tends to be larger $\ge$ 20 kpc above the disk in TORH than in Hydro.  Thus, throughout most of the tail, including a magnetic field increases the tail velocity, making more likely that stripped gas will leave the TORH box.  However, within 20 kpc of the disk, the wind accelerates gas the most quickly in the Hydro run and there tends to be more gas within 15 kpc with negative z-velocities in TORH and DIP than in Hydro.  If anything, Figures \ref{fig-diskmass} and \ref{fig-tailzvz} indicate that TORH has more fallback from 170 Myr to 310 Myr than Hydro.  We find that the MHD runs will be at least as likely as the Hydro run to remove gas from the tail through fallback near the disk or acceleration out of the simulated domain.  

\begin{figure*}
\begin{center}
\includegraphics[scale=1.1,trim= 0mm 0mm 19mm 0mm, clip]{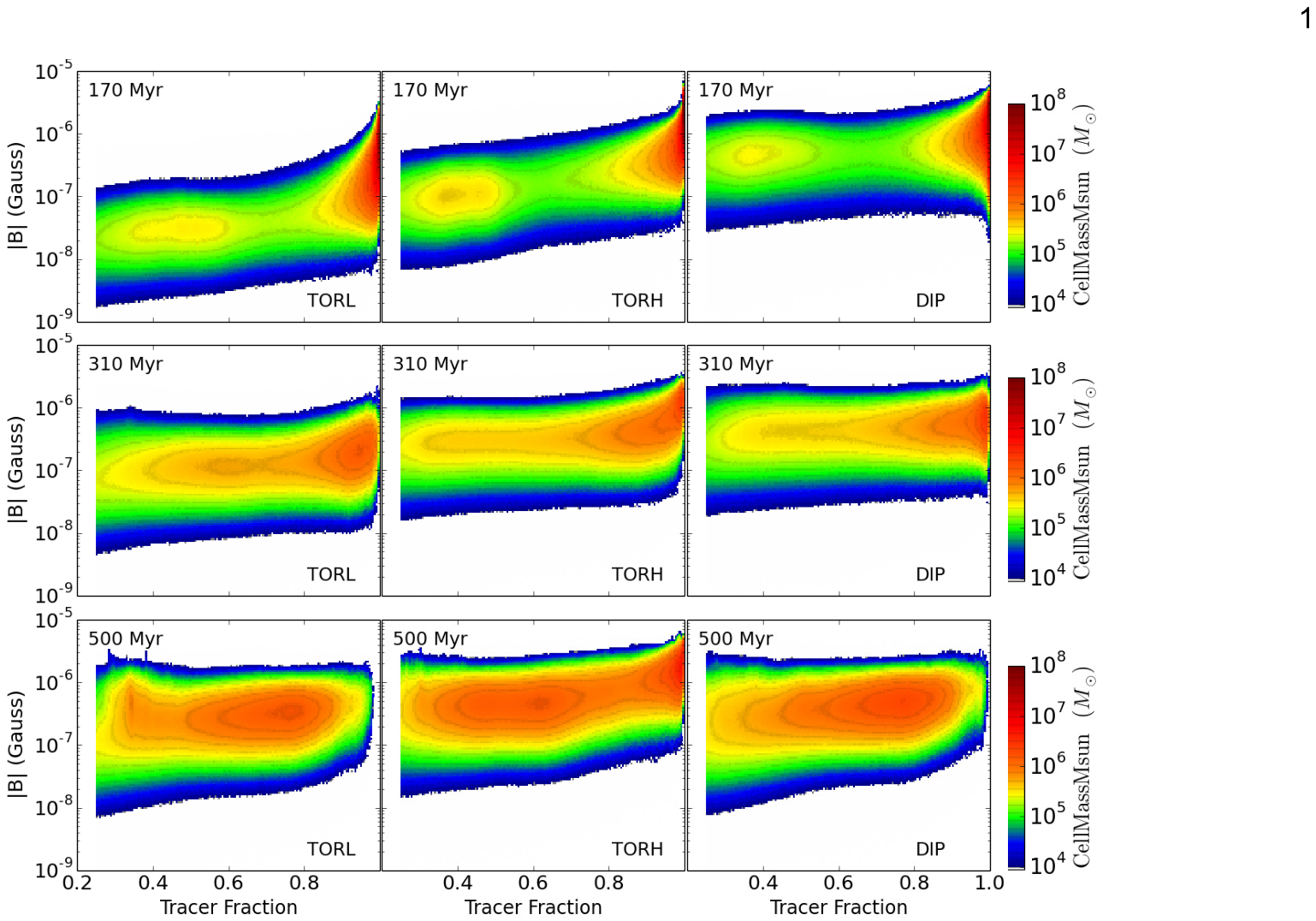}

\caption{\scriptsize{Mass contours of tail gas as a function of tracer fraction and B magnitude.  Tail gas is defined as in Figure \ref{fig-tailzvz}.  The columns from left to right are TORL, TORH, DIP.  At later times it become clear that there is more unmixed gas in the TORH tail than in either of the other tails. } }\label{fig-tailssB}
\end{center}
\end{figure*}

Therefore, it is worthwhile to examine more directly whether the magnetic field impedes gas mixing.  To do this, we plot the tail gas mass as a function of magnetic field magnitude and tracer fraction in Figure \ref{fig-tailssB}.  170 Myr after the wind has hit the disk the initial, fast stripping has just ended in all three runs, so the gas in the tail was relatively recently removed from the disk and very little gas originating in the disk has left the simulation box.  In the top panels of Figure \ref{fig-tailssB}, gas with the highest magnetic field magnitude is also the least mixed.  At least part of this is because gas from the more central regions of the disk, with stronger magnetic fields, is stripped at later times.  This interpretation can explain the differences that we see between the top panels of the three runs:  TORL and TORH have the same magnetic field structure in their disks and in both tails the magnetic field strength drops by a factor of about 6 from the unmixed maximum contour level to the mixed maximum contour at a tracer fraction of about 0.4.  However, the magnetic field in DIP only drops by about a factor of two from the unmixed to the mixed peak, and the magnetic field strength is much more constant throughout the disk in DIP than in the toroidal field runs (Figure \ref{fig-bfield}).     

Recall that from one output to the next, the gas we are examining is changing.  We expect that gas that is recently stripped has a high tracer fraction, then at later times will have a lower tracer fraction as it mixes with the surrounding ICM (moving to the left in the panels in Figure \ref{fig-tailssB}).  At 310 Myr, more gas in TORL has a tracer fraction of 0.6 than in the TORH or DIP runs, which have most of their tail gas at higher tracer fractions.  At 500 Myr, TORH clearly has more gas residing at higher tracer fractions than either TORL or DIP.  As we know from Figure \ref{fig-diskmass}, the stripping rate starting at about 310 Myr is very similar across all the simulations, so the stripped gas from 310 Myr to 500 Myr has had a similar time in which to mix with the ICM.  This is a strong indicator that gas mixes more slowly in the strongest magnetic field case, and the initial field morphology in the disk has little affect on the tail.
  
\begin{figure*}
\begin{center}
\includegraphics[scale=1.1,trim= 0mm 0mm 19mm 0mm, clip]{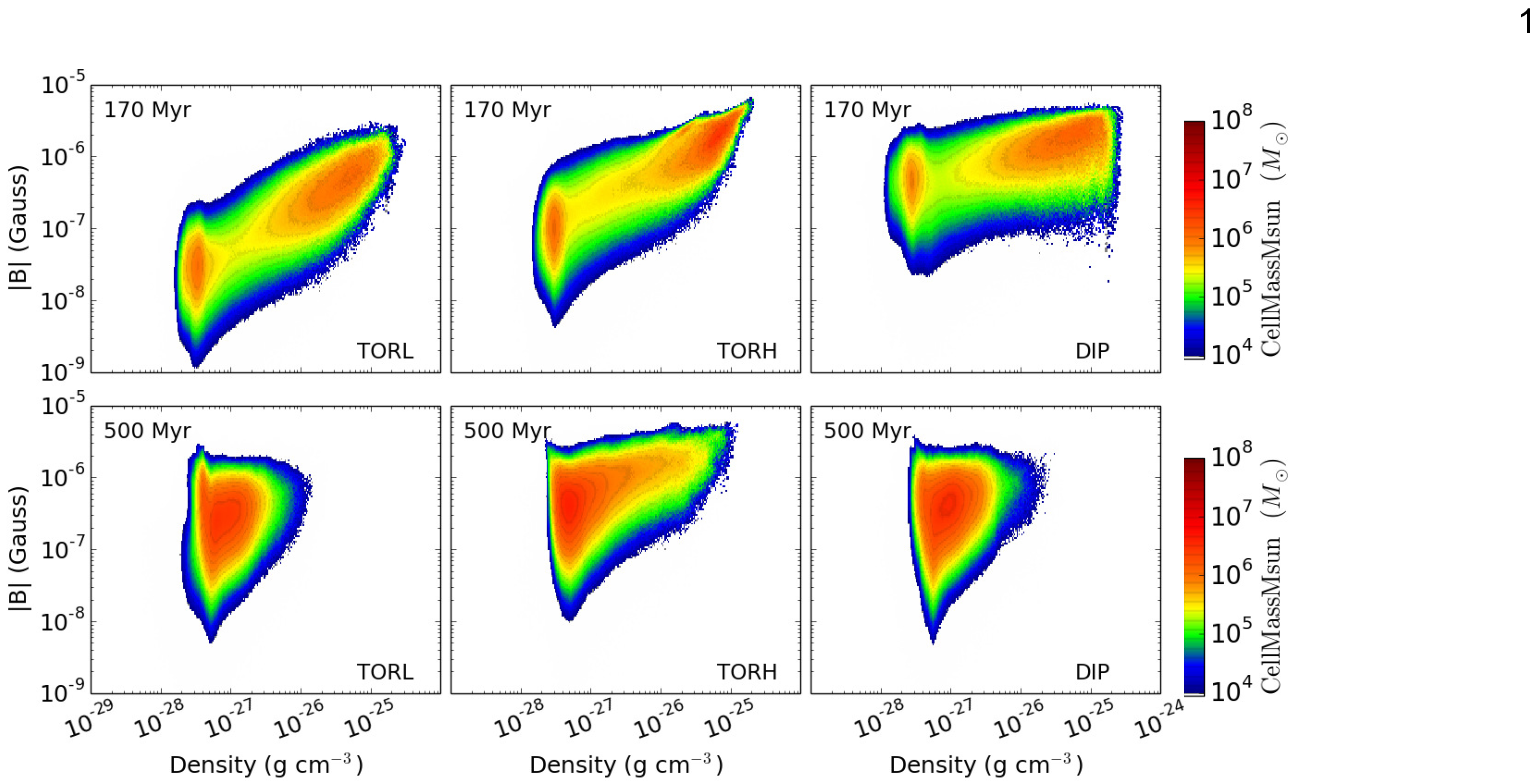}

\caption{\scriptsize{Mass contours of tail gas as a function of density and B magnitude.  Tail gas is defined as in Figure \ref{fig-tailzvz}.  The columns from left to right are TORL, TORH, DIP.  TORH, with the strongest magnetic field, has much more high-density gas at 500 Myr than the other two MHD runs.} }\label{fig-tailrhoB}
\end{center}
\end{figure*}

In Figure \ref{fig-tailrhoB} we also consider this problem by examining observable gas properties, the density and magnetic field strength.  At early times, 170 Myr after the wind has hit the disk, all three runs have a similar range and distribution of gas density.  By 500 Myr after the wind has hit the disk, only the TORH run has gas in the tail with densities greater than 3$\times$10$^{-26}$ g cm$^{-3}$.  Tonnesen \& Bryan (2010) find that high density gas in the tail moves more slowly, so this gas is the least likely to have left the simulated box.  Further, fallback is also dominated by lower density gas that can be more easily pushed into the shadow of the disk by disordered motion (Figure \ref{fig-diskrhovz} and Tonnesen \& Bryan 2010).  Therefore, it is likely that in TORL and DIP the high density gas has mixed with the lower density ICM and that the stronger magnetic field in TORH inhibits mixing.

It is important to note that we do not include explicit diffusion in our simulation.  This affects gas mixing and sets our magnetic Prandtl number to 1.  We highlight that because we do not include diffusion, mixing between magnetized galactic gas and unmagnetized intracluster gas can only occur on scales smaller than our cell size.  See Ruszkowski et al. (2014) for a detailed discussion of mixing in ideal MHD simulations.  

A magnetic Prandtl number of one means that the viscous dissipation length is the same as the resistive dissipation length, so our velocity structures will be the same size as our magnetic field structures.  
Since $\eta$ $<<$ $\nu$ 
in intracluster plasma (e.g. Brandenburg
\& Subramanian 2005), the ICM has large Pm (as does the ISM).  Simulations that vary the magnetic Prandtl number indicate that this difference could effect the magnetic field in our tails.  For example, Fromang et al. (2010) found that MRI turbulence is not sustained when Pm$\le$1.  Also, at larger Pm, turbulent flows may produce more magnetic energy, closer to equipartition with kinetic energy (Subramanian et al. 2006 and references therein), although values of Pm as large as those expected in intracluster gas have not been simulated.  Bovino et al. (2013) find that at the large Pm and Rm values expected in the ICM the turbulent growth rate is much larger than at Pm$\sim$1.  Therefore, our Pm=1 may result in weaker magnetic fields in our stripped tails and faster decay of turbulence than we would expect in observed tails evolving in a high-Pm ICM.  

\section{Magnetizing the ICM}\label{sec:BICM}

Thus far we have focused on how a galactic magnetic field will affect the gas in the disk and tail.  However, the magnetic field in the tail should be examined in its own right, as it may help to magnetize the ICM. 

\subsection{Growth of Magnetic Field in the Stripped Tail}

\begin{figure}[!h]
\begin{center}
\includegraphics[scale=1.08,trim= 0mm 0mm 5mm 0mm, clip]{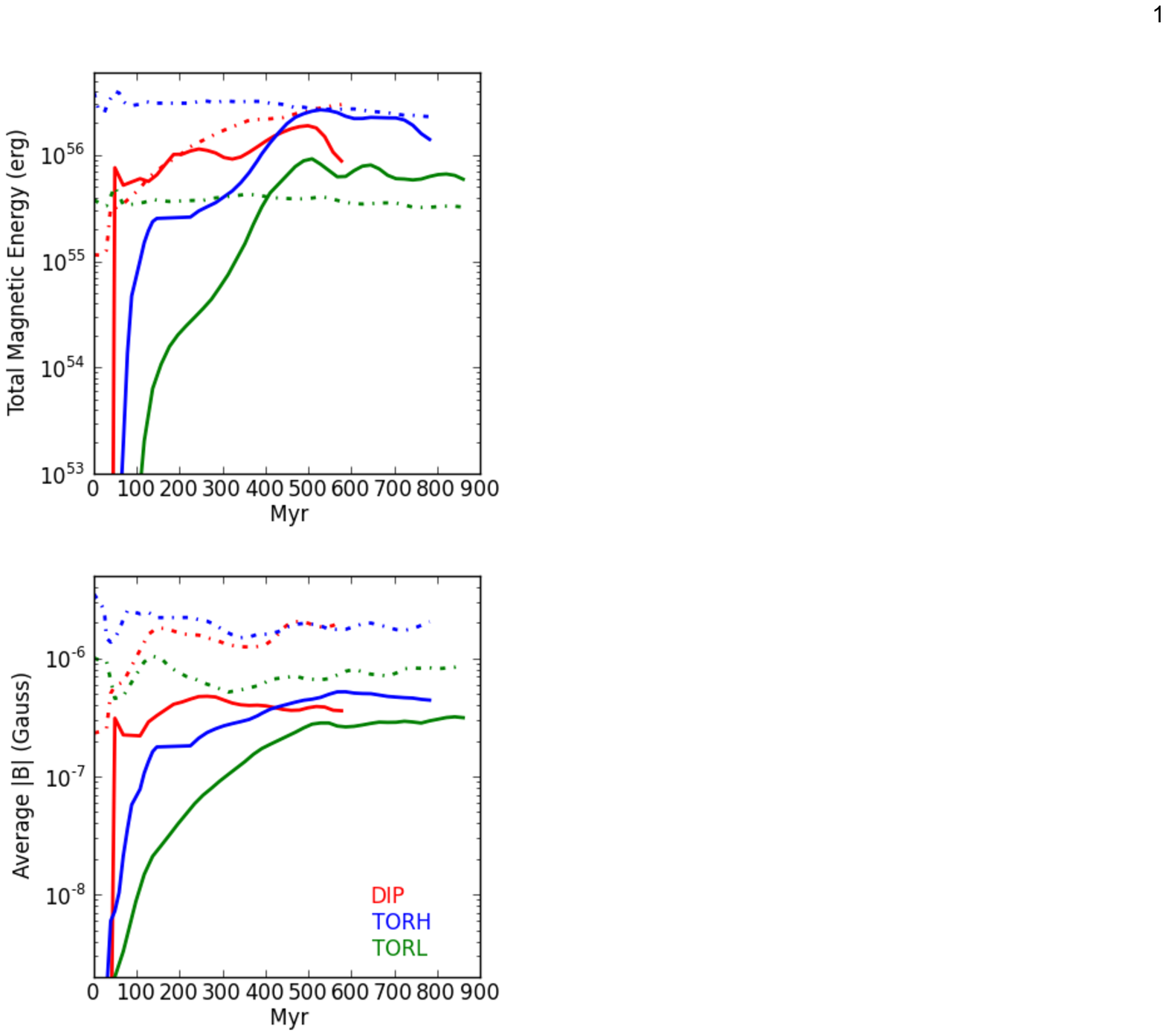}\\
\caption{\scriptsize{\textbf{Top Panel:}  The total magnetic energy in the disk (dot-dashed lines) and tail (solid lines) for the three MHD runs.  There is a dramatic increase in the magnetic energy in the tails over time, and the tails add magnetic energy to the ICM. \textbf{Bottom Panel:}  The average magnetic field in a cell in the disk (dot-dashed lines) and tail (solid lines) for the three MHD runs.  }}\label{fig-Benergy}
\end{center}
\end{figure}

In Figure \ref{fig-Benergy} we plot the total magnetic energy (top panel) and average magnetic field strength (bottom panel) in the tail and in the disk.  The DIP run is in red, the TORH run in blue, and the TORL run in green.  The solid lines denote the values in the tail (more than 10 kpc above the central plane of the galaxy) and the dash-dotted lines show the values in the disk (a cylindrical region with h = 10 kpc and r = 28.6 kpc).  The total magnetic energy is the sum of B$^2$/$8\pi$ in each cell times the cell volume.  

Focusing first on the disk magnetic energy, we find that both of the galaxies with a toroidal field have relatively constant magnetic energy in the disks.  This is not surprising as the strongest magnetic fields reside within the stripping radius of the disk ($\sim$15 kpc).  We also see that the magnetic energy density in the disk in the DIP run increases with time.  As we discussed in Section \ref{sec:magfield}, this is because the magnetic field is not in a steady state and a toroidal component grows with time due to radially-varying velocity.

The total magnetic energy in the tail in all three simulations generally increases with time, particularly for the first 500 Myr.  This is quite interesting as we have made no correction for material leaving the box in this figure.  Material stripped from the disk will continue to add gas with stronger magnetic fields, and turbulent stretching of the magnetic field will increase the total magnetic energy in the tail.  Indeed, in our weakest MHD run, TORL, the magnetic energy in the tail is larger than the magnetic energy in the disk later than about 400 Myr after the wind has hit the disk.  

However, we note that as shown in the bottom panel of Figure \ref{fig-Benergy}, the mean magnetic field strength in the tail is always less than in the disk.  Again, this is because the magnetic field strength increases towards the center of the disk, where ram pressure is not strong enough to remove the disk gas in our simulations.  The mean magnetic field in our tails is between 0.3-0.6 $\mu$G, which is within a factor of a few of the observationally-determined intracluster magnetic field strengths outside of cluster centers. 

\subsection{MicroGauss Fields in the ICM}

\begin{figure*}
\begin{center}
\includegraphics[scale=1.05,trim= 0mm 0mm 6mm 0mm, clip]{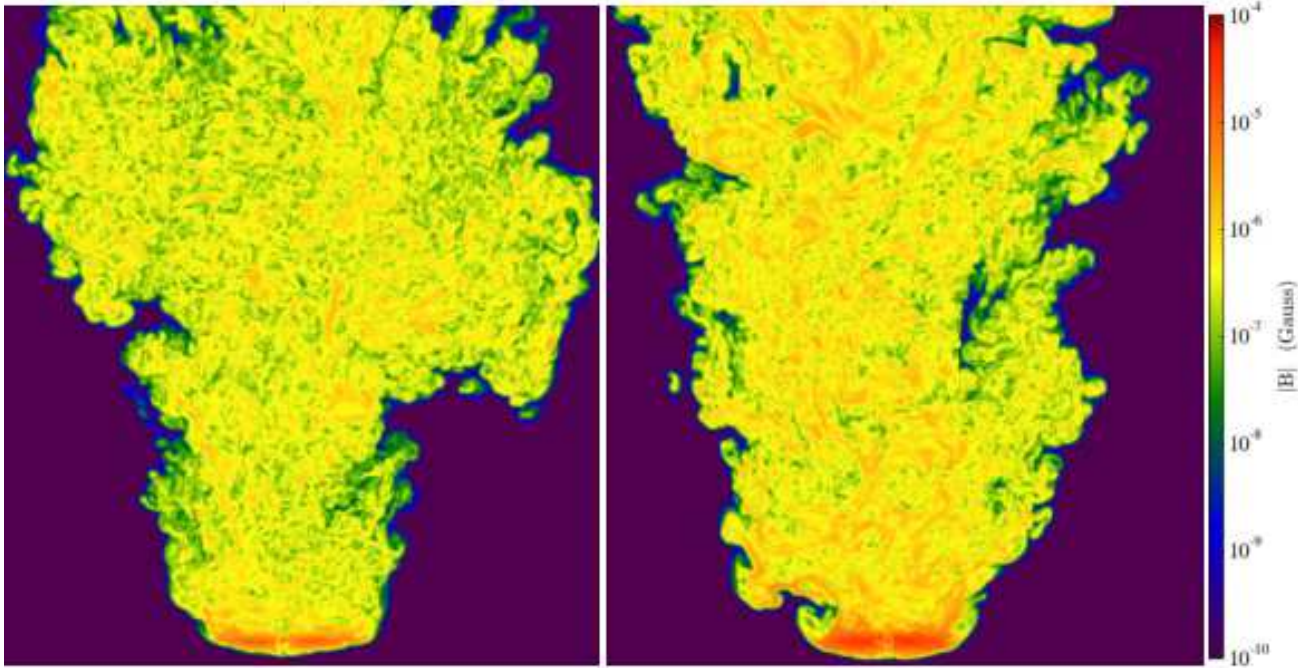}
\caption{\scriptsize{Slices of the magnetic field strength 750 Myr after the wind has hit the disk in the TORL (left panel) and TORH (right panel) runs.  We chose these two runs because they have the lowest (TORL) and highest (TORH) amount of gas with magnetic field strengths of at least a $\mu$G. The image region is 94 $\times$ 102 kpc.  The right panel can be compared to the top panel in Figure \ref{fig-mixing}.}}\label{fig-Bslices}
\end{center}
\end{figure*}

\begin{figure}[!h]
\begin{center}
\includegraphics[scale=0.74,trim= 0mm 0mm 0mm 0mm, clip]{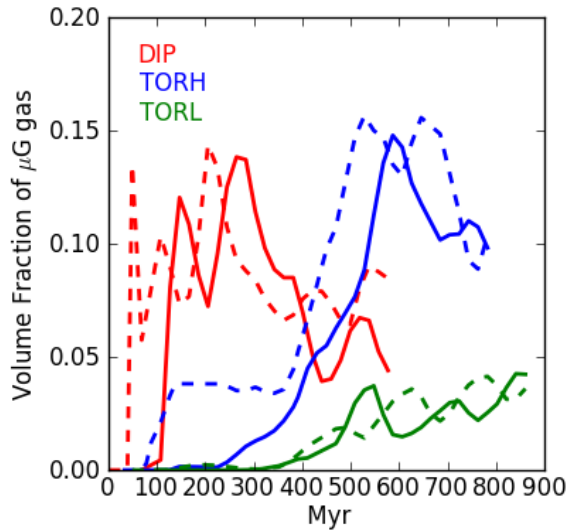}
\caption{\scriptsize{The volume fraction of gas with $\ge$$\mu$G magnetic fields in tails in the DIP (red), TORH (blue), and TORL (green) simulations.  The dashed lines show the fraction from 10-50 kpc above the disk, and the solid lines show the fraction from 50-100 kpc.  The volume fraction is similar, so the total volume of $\mu$G gas increases as we move farther along the tail, although sometimes with a time lag between the near and more distant tail regions.  }}\label{fig-Bvol}
\end{center}
\end{figure}  

In this section we consider whether magnetic fields of $\ge$$\mu$G strength exist in our stripped tails, as several observations indicate that intracluster magnetic fields are $\sim$$\mu$G (e.g. Bonafede et al. 2009; Clarke, Kronberg \& B{\"o}hringer 2001).  From Figures \ref{fig-tailssB} \& \ref{fig-tailrhoB} we know that $\mu$G fields do exist in our tail.  In Figure \ref{fig-Bslices} we show slices of the magnetic field strength 750 Myr after the wind has hit the disk in the TORH and TORL runs, which have the strongest (TORH) and weakest (TORL) magnetic fields in the tail.  The structure of the B field in the TORH slice resembles the structure in our mixing slice in Figure \ref{fig-mixing}.  This is as we would expect for turbulence to be driving both the increase in our magnetic energy density and mixing between stripped gas and the ICM.  

From a visual inspection of Figure \ref{fig-Bslices} it is clear that the correlation length of the field in the tail is smaller than galaxy scales.  This is somewhat smaller than the scale indicated by Faraday rotation observations, which imply lengths of 10-20 kpc (e.g. Clarke et al. 2001; Eilek \& Owen 2002; Clarke 2004).  Using a power-law fit to the B-field observed with Faraday rotation measure and fractional polarization images of the radio galaxy in A2199 Vacca et al. (2012) find the best fit for the maximum scale of magnetic fluctuations to be 35 kpc.  However, Vogt \& En{\ss}lin (2003, 2005) find correlation lengths of less than 5 kpc in three clusters and generally argue that the magnetic field correlation length is in fact 2-4 times shorter than the rotation measure fluctuation scale with which it is often equated.  In a re-examination of one of those clusters, Kuchar \& En{\ss}lin (2011) argue that there is magnetic power up to at least 8 kpc length scales.  To directly compare our magnetic field length scales to observations we will need to perform mock observations on a much larger volume of a cluster or observations would need to pinpoint stripped tails in clusters.

In Figure \ref{fig-tailzvz}, the width of the orange contour indicates that the turbulent velocity is a few hundred km/s.  The turbulent velocities in the x and y directions seem to be similar based on slices of the x-, y-, and z- velocity.  Thus, for the lower-density gas in our tail, 4 $\times$10$^{-28}$ g cm$^{-3}$, and a conservative estimate of the turbulent velocity magnitude of $\sqrt{3}$$\times$100 km/s, the average magnetic field strength through equipartition should be about 1 $\mu$G.  This is a factor of 2-3 higher than we find in the tail (bottom panel of Figure \ref{fig-Benergy}), so equipartition is either not reached in this system, or would only be reached after the gas has left our simulated region.

In Figure \ref{fig-Bvol}, we plot the volume fraction of gas with $\ge$$\mu$G magnetic fields in the tails in the DIP (red), TORH (blue), and TORL (green) simulations.  The dashed lines show the fraction from 10-50 kpc above the disk, and the solid lines show the fraction from 50-100 kpc.  The volume fraction is similar, so the magnetic field strength does \textit{not} fade with distance from the disk.  Also, the total volume of $\ge$$\mu$G gas increases as we move farther along the tail, although sometimes with a time lag between the near and more distant tail regions.  

We clearly see that the volume fraction of $\ge$$\mu$G gas depends strongly on the galactic magnetic field strength by comparing the fraction in TORH and TORL, so either the turbulence is much stronger in TORH than in TORL (which does not seem to be the case looking at the velocity structure in Fig. \ref{fig-tailzvz}), or the seed field from the disk is more important than a turbulent dynamo in determining the magnetic field strength in the tail.  

\section{Discussion}\label{sec:discussion}

\subsection{Comparison with previous work}\label{sec:simcomp}

We compare our results to earlier work that has examined similar properties of ram pressure stripped disks and tails.  First, we find that the remaining disk gas mass and radius in our simulations agree well with the non-cooling hydrodynamic runs in Tonnesen \& Bryan (2009) and  Roediger \& Br{\"u}ggen (2006).  The differences between our results and theirs are because we use the tracer fraction rather than a density cut to define our disk gas and because we use the maximum radius at which high tracer fraction gas can be found to define our radius (rather than the minimum radius at which low density gas can be found).  Also, our tail is very similar to the non-cooling tail in Tonnesen \& Bryan (2010), morphologically, in density-temperature space, and in velocity structure.  The range of resolutions used,  40-80 pc in Tonnesen \& Bryan (2009; 2010), 159 pc in this work, and 500 pc in Roediger \& Br{\"u}ggen (2006) highlights that these results are insensitive to resolution.  

Ruszkowski et al. (2014) examined the impact of intracluster magnetic fields on stripping of disk galaxies, but did not include galactic magnetic fields.  Because they use a different galaxy model, we cannot compare our results directly to theirs and can only make qualitative comparisons.  They find that including a magnetic field in the ICM changes the stripping rate from that of the pure hydrodynamic case with a face-on wind.  The initial stripping rate is the same (for the first $\sim$100 Myr), then the MHD run has slower stripping to a maximum difference in the disk gas of less than 15\%, at which point the MHD stripping rate increases such that $\sim$550 Myr after the wind has hit the disk the MHD and hydrodynamical runs have the same amount of gas in the disk.  The authors posit that the slower stripping for $\sim$400 Myr  is due to the magnetic draping layer that forms on, and protects, the face of the galaxy.  This is a process quite separate from any that we simulate, so there is no tension between the two results.  

Unlike Ruszkowski et al. (2014), we do not find a dramatic difference in the morphology of our tails with and without magnetic fields.  The differences observed in Ruszkowski et al. (2014) may occur because they also include radiative cooling, a process that dramatically affects tail structure, as discussed in detail in Tonnesen \& Bryan (2010).  

This paper compares, in hydrodynamic and magnetohydrodynamic runs, the morphology of gas in the tail, the $\rho$-T distribution of gas in the tail, and examines how stripped gas may be mixed into the ICM, all without including radiative cooling.  However, Tonnesen \& Bryan (2010; also Tonnesen et al. 2011) have stressed the necessity of including radiative cooling in order to reproduce HI, H$\alpha$, and X-ray observations of ram pressure stripped tails.  Including radiative cooling, however, can result in disks and tails that are clumpier than those observed (Tonnesen \& Bryan 2009; Ruszkowski et al. 2014) and may miss the more diffuse ISM that would become diffuse stripped gas.  While, as these authors have shown, this low-density gas will not be observed in HI and H$\alpha$ emission, understanding how this gas mixes with the ICM is very important to understanding how ram pressure stripping will pollute the ICM with metals and magnetic fields.  Our examination of how magnetic fields affect the diffuse gas tail is a step in understanding that process.     

In this paper, we only consider a face-on wind geometry.  The role of the inclination angle in disk gas stripping has been considered in previous work (e.g. Quilis et al. 2000; Vollmer et al. 2001; Schulz \& Struck 2001; Roediger \& Br{\" u}ggen 2006; J{\' a}chym et al. 2009), so here we do not focus on tilted disks.  
However, we can briefly discuss the possible impact of galactic magnetic fields on the ram pressure stripping of tilted disks.  The previous works listed above have generally found that inclination angle does not have a strong impact on the amount of gas stripped from a disk until the wind is close to edge-on, at which point much less gas is removed from the galaxy.  We see no reason for the inclusion of galactic magnetic fields to change these results.   
As we have found, the stripping rate of gas that is bound by magnetic fields does not differ from the stripping rate of the purely hydrodynamical case, so we would not expect this result to change with galaxy inclination.  However, as we have discussed, the disk in TORH has expanded in the z-direction due to the strong magnetic pressure.  In the highly inclined case, we would expect more gas removal in the TORH run because the more distant gas has a weaker gravitational restoring force from the disk.  We expect that at high inclination angles the height of the disk gas above the galaxy plane is more important than whether there are magnetic fields.

\subsection{Comparison With Observations}\label{sec:obs}

As we discussed in our introduction, Murphy et al. (2009) compare maps of the FIR-radio correlation between ram pressure stripped and normal galaxies, and find radio deficits along the face of the interaction between the galaxy and the ICM.  They believe that these radio deficits are due to the sweeping out of low density gas and the corresponding magnetic fields rather than compression towards the disk plane because there is no ridge of strong radio emission between the radio deficit and the galaxy mid plane.  We look for physical insight into these observations by examining the magnetic field magnitude of disk gas as a function of distance below the disk.  We can only look at the TORL and TORH runs, because in the DIP run the differential rotation in the disk increases the central magnetic field strength, leaving us unable to determine whether any of the increase in the magnetic field strength is due to compression from the ICM wind.  

\begin{figure*}
\begin{center}
\includegraphics[scale=1.1,trim= 0mm 0mm 10mm 0mm, clip]{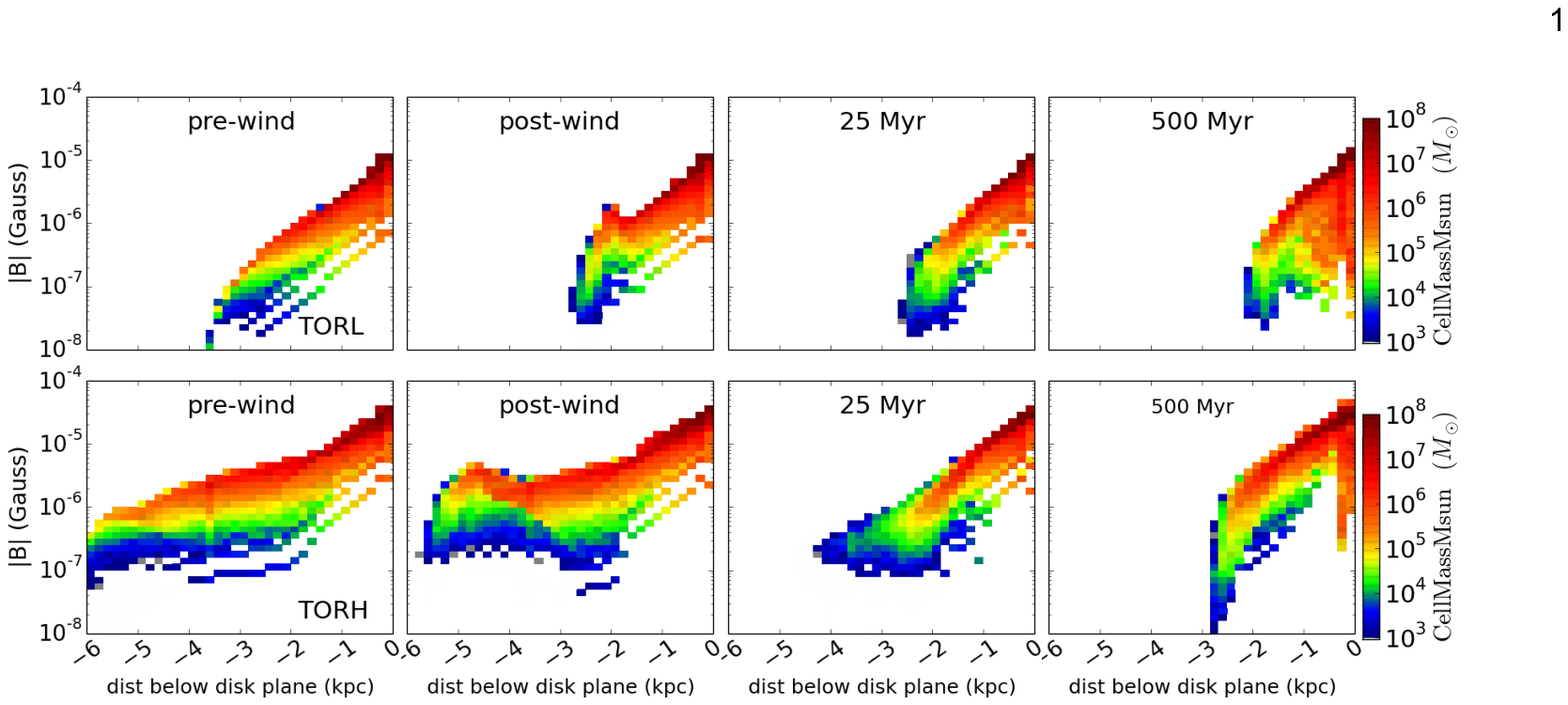}

\caption{\scriptsize{Mass contours of central disk gas that lies below the disk mid plane as a function of B Magnitude and distance from the disk plane.  Here central disk gas is defined as gas within the central 5 kpc radius with a tracer fraction of more than 0.6.  The top row is TORL and the bottom is TORH.  Each column is a different time step:  t=25 Myr, the output immediately before the wind hits the disk, t=30 Myr, the output at which the wind hits the disk (reaches the mid plane of the disk), t=55 Myr, or 25 Myr after the wind has hit the disk, and t=530 Myr, or 500 Myr after the wind has hit the disk.  As the wind initially hits the disk the B magnitude increases as a compression wave moves across the disk.  However, by 25 Myr after the wind hits the disk, there is no increase in the B magnitude anywhere on the wind-facing side of the central disk.  In the last column we see that this remains true throughout the simulations.}}\label{fig-diskzB}
\end{center}
\end{figure*}

In Figure \ref{fig-diskzB}, we plot contours of the amount of gas mass within a 5 kpc radius cylinder around the disk center as a function of magnetic field magnitude and z-distance below the disk plane.  The range of $|$B$|$ magnitude at a single distance below the disk is due to the range in galactic radius.  We choose to focus on the small central region so that the figure does not become more complicated by a larger range of magnetic field values.  We show the output before the wind hits the disk, the output at which the wind hits the disk (the shock front in the ICM has just passed the disk plane), 25 Myr after the wind has hit the disk, and 500 Myr after the wind has hit the disk.  

Immediately as the wind hits the disk, we see that the magnetic field strength increases as a shock propagates through the disk.  This increase in $|$B$|$ magnitude corresponds to an increase in density from the shock.  However, this increase is short-lived, and by 25 Myr after the wind has hit the disk there is no enhancement of the magnetic field (or density).  As predicted in Murphy et al. (2009) and Vollmer et al. (2010), lower density gas is swept away.  The edges continue to be ablated, so gas that is not quickly swept away mixes with the non-rotating ICM and is eventually removed, as can be seen by the shrinking of the disk in the z-direction from 25 Myr to 500 Myr after the wind has hit the disk.  

\section{Conclusions}\label{sec:conclusion}

We have run high-resolution galaxy simulations including galactic magnetic fields in order to understand how galactic magnetic fields affect ram pressure stripping and mixing into the ICM.  We compare a hydrodynamical simulation to two simulations with toroidal galactic magnetic fields (TORL and TORH) and a run with a dipole-like magnetic field (DIP).  Our main conclusions are:

1.  Magnetic fields in the galactic disk do not dramatically change the stripping rate or amount of gas removed from Milky Way-type galaxies.  Even including a field with a radial component that links high and low-density gas does not effect the initial stripping rate due to magnetic tension dragging gas from the disk at the field strength we simulate.  The stripping profiles in all four runs are nearly identical beyond 360 Myr after the wind has hit the disk (Figure \ref{fig-diskmass}).

2.  The density and pressure of gas in the disk is also very similar in all four runs, although we see some evidence that a magnetic field inhibits the acceleration of stripped gas at later times.  Specifically, we find that in the TORH and DIP runs the bulk of the gas being removed by the ICM wind leaves the disk with lower velocities than in the TORL and Hydro runs (Figures \ref{fig-diskrhovz} \& \ref{fig-diskpvz}).     

3.  The velocity structure in the tail is very similar in all four runs, with two consistent differences: first, within 20 kpc of the disk, the Hydro run has more gas accelerated to high velocities than any MHD run, and second, more than 20 kpc above the disk, the majority of gas in the tail in the TORH and DIP runs is moving away from the disk at velocities equal to or greater than the tail gas in the Hydro run (Figure \ref{fig-tailzvz}).

4. More dense gas survives in the MHD tails from 170 Myr to 310 Myr (Figure \ref{fig-tailrhoT}).  By 500 Myr, the difference between the MHD and Hydro runs has shrunk, but persists throughout the simulations.  We use Figures \ref{fig-tailssB} and \ref{fig-tailrhoB} to determine that this is because the magnetic field in the stripped tail inhibits mixing with the surrounding ICM.  

5.  The magnetic energy in the tail increases with time as the volume of the tail increases (Figure \ref{fig-Benergy}).  We find that the mean magnetic field in the tail seems to plateau between 0.3-0.6 $\mu$G, and up to 15\% of the volume of the tail has magnetic field strengths of at least 1 $\mu$G, but this depends on the strength of the magnetic field in the disk.  Indeed, the field strength in the tail seems to depend more on the galactic magnetic field strength than on turbulent enhancement.

6.  We examine the magnetic field on the wind-facing side of our disk and find that the magnetic field only briefly ($\sim$25 Myr) increases due to compression from the shock front traveling through the disk, and otherwise the magnetic field strength is deficient in comparison to the field strength before the wind hits.  This is in good agreement with the observational findings of Murphy et al. (2009), and does not require ram pressure to affect the star formation rate in galaxies.  

Although the tails that we are modeling are disordered flows that vary with time, we find that including magnetic fields will allow mostly unmixed gas to survive to larger distances from the disk-both by inhibiting mixing and by allowing for more acceleration of the tail gas by the ICM wind.  We also find initial evidence that ram pressure stripping can magnetize the ICM.  While only 5-15\% of the gas has $\mu$G magnetic field, we see from Figure \ref{fig-Bslices} that a much larger fraction of the tail has 0.1 $\mu$G field strengths, and that the mean field in the tail is at least 0.3 $\mu$G, which is within a factor of a few of the observationally-inferred intracluster magnetic field strength outside the cluster core.  As we discuss (\S \ref{sec:rhoTtail}), because our magnetic Prandtl number of 1 is orders of magnitude below Pm in the ICM and ISM, turbulence in our simulated tails likely strengthens the magnetic field much less than in nature.

It is also worth noting that our ram pressure strength and ICM velocity were selected from a sample of galaxy orbits in a cosmologically-simulated cluster (with M$_{200}$$\sim$4$\times$10$^{14}$ M$_{\odot}$) taken from about the virial radius (Tonnesen et al. 2007; Tonnesen \& Bryan 2009).  As shown in Figure \ref{fig-diskmass}, these galaxies are stripped of less than 60\% of their gas, leaving their inner disks intact.  As galaxies experience higher ram pressure, particularly closer to the cluster center, they may be completely stripped, polluting the ICM with stronger magnetic fields. 

Our tails are at least 100 kpc long and more than 30 kpc wide, so could contribute to a large area-covering fraction in a cluster.  Assuming no overlap, there would need to be about 165 tails that are 30 kpc in diameter and 200 kpc long within 1 Mpc for an area covering factor of unity.  Solanes et al. (2001) find 171 HI deficient galaxies within 1 Abell radius (1.5 $h^{-1}$ Mpc) in the Virgo cluster.  If all of these galaxies have been ram pressure stripped with $\sim$200 kpc tails, similar to but longer than we are able to simulate in our small domain, magnetic fields in stripped tails would have an area-covering factor of $\sim$25\% with 2 Mpc of the Virgo cluster center.  This may be a lower estimate of the number of galaxies that can contribute to the intracluster magnetic field, because if magnetic fields thread through galactic gas halos, galaxies that are not HI deficient could also contribute magnetic energy to the ICM.  We can test if strong stripping is necessary to add significant magnetic energy to the ICM by simply rerunning our simulations with slower and lower-density winds.  

In the future we will simulate a larger box to determine how long unmixed gas survives in the ICM, predicting how long strong metallicity gradients will survive in cluster gas.  While in this work we are using ideal MHD, we also plan to include anisotropic conduction, which may affect the heating and mixing of the tail gas relative to that in the hydrodynamic simulation.  Determining tail lengths and magnetic field strength and correlation length in the tails is important for determining how much of the magnetic field measured in clusters could come from ram pressure stripped galaxies.  Future observational constraints on the magnitude, volume filling factor and correlation length of the intracluster magnetic field will allow us to more clearly determine the fraction of the total intracluster magnetic field that can be attributed to magnetized stripped tails.

\vskip 1cm
We acknowledge support from NSF grant PHY-1144374 and from Princeton University through the Lyman Spitzer, Jr. Fellowship.  Computations were performed on resources provided by the Princeton Institute for Computational Science and Engineering.  We thank Professor Greg Bryan for useful discussions, and Professor Ming Sun and Professor Jeffrey Kenney for useful comments.  We also thank our anonymous referee for comments and suggestions that improved the quality of this paper.

\end{document}